\documentclass[preprint2]{aastex}

\newcommand{\myemail}{trawle@as.arizona.edu}

\shorttitle{``Warm dust" galaxies in clusters at z$\sim$0.3}
\shortauthors{Rawle et al.}

\begin{document}

\title{Discovery of ``warm dust" galaxies in clusters at z$\sim$0.3: evidence for stripping of cool dust in the dense environment?\footnote{Partially based on data from {\it Herschel}, an ESA space observatory with science instruments provided by European-led Principal Investigator consortia and with important participation from NASA.}}

\author{T.~D.~Rawle\altaffilmark{1}, M.~Rex\altaffilmark{1}, E.~Egami\altaffilmark{1}, S.~M.~Chung\altaffilmark{2,3}, P.~G.~P\'{e}rez-Gonz\'{a}lez\altaffilmark{4,5}, I. Smail\altaffilmark{6}, G.~Walth\altaffilmark{1}, B.~Altieri\altaffilmark{7}, P.~Appleton\altaffilmark{8}, A.~Berciano~Alba\altaffilmark{9,10}, A.~W.~Blain\altaffilmark{11}, M.~Dessauges-Zavadsky\altaffilmark{12}, D.~Fadda\altaffilmark{8}, A.~H.~Gonzalez\altaffilmark{2}, M.~J.~Pereira\altaffilmark{1}, I.~Valtchanov\altaffilmark{7}, P.~P.~van~der~Werf\altaffilmark{10}, M.~Zemcov\altaffilmark{13,14}}

\email{\myemail}

\altaffiltext{1}{Steward Observatory, University of Arizona, 933 N. Cherry Ave, Tucson, AZ 85721}
\altaffiltext{2}{Department of Astronomy, University of Florida, Gainesville, FL 32611-2055, USA}
\altaffiltext{3}{Department of Astronomy, The Ohio State University, 140 West 18th Avenue, Columbus, OH 43210, USA}
\altaffiltext{4}{Departamento de Astrof\'{\i}sica, Facultad de CC. F\'{\i}sicas,Universidad Complutense de Madrid, E-28040 Madrid, Spain}
\altaffiltext{5}{Associate Astronomer at Steward Observatory, University of Arizona}
\altaffiltext{6}{Institute for Computational Cosmology, Durham University, South Road, Durham DH1 3LE}
\altaffiltext{7}{{\it Herschel} Science Centre, ESAC, ESA, PO Box 78, Villanueva de la Ca\~{n}ada, 28691 Madrid, Spain}
\altaffiltext{8}{IPAC, California Institute of Technology, Pasadena, CA 91125}
\altaffiltext{9}{ASTRON, Oude Hoogeveensedijk 4, 7991 PD Dwingeloo, the Netherlands}
\altaffiltext{10}{Sterrewacht Leiden, Leiden University, PO Box 9513, 2300 RA, Leiden, the Netherlands}
\altaffiltext{11}{Department of Physics \& Astronomy, University of Leicester, University Road, Leicester LE1 7RH, UK}
\altaffiltext{12}{Observatoire de Gen\`{e}ve, Universit\'{e} de Gen\`{e}ve, 51 Ch. des Maillettes, 1290, Sauverny, Switzerland}
\altaffiltext{13}{California Institute of Technology, Pasadena, CA 91125}
\altaffiltext{14}{Jet Propulsion Laboratory, Pasadena, CA 91109}


\begin{abstract}

Using far-infrared imaging from the ``{\it Herschel} Lensing Survey'', we derive dust properties of spectroscopically-confirmed cluster member galaxies within two massive systems at z$\sim$0.3: the merging Bullet Cluster and the more relaxed MS2137.3-2353. Most star-forming cluster sources ($\sim$90\%) have characteristic dust temperatures similar to local field galaxies of comparable infrared (IR) luminosity ($T_{\rm dust}$ $\sim$ 30 K). Several sub-LIRG ($L_{\rm IR}$ $<$ 10$^{11}$ L$_{\odot}$) Bullet Cluster members are much warmer ($T_{\rm dust}$ $>$ 37 K) with far-infrared spectral energy distribution (SED) shapes resembling LIRG-type local templates. X-ray and mid-infrared data suggest that obscured active galactic nuclei do not contribute significantly to the infrared flux of these ``warm dust" galaxies. Sources of comparable IR-luminosity and dust temperature are not observed in the relaxed cluster MS2137, although the significance is too low to speculate on an origin involving recent cluster merging. ``Warm dust" galaxies are, however, statistically rarer in field samples ($>$ 3$\,\sigma$), indicating that the responsible mechanism may relate to the dense environment. The spatial distribution of these sources is similar to the whole far-infrared bright population, i.e. preferentially located in the cluster periphery, although the galaxy hosts tend towards lower stellar masses (M$_*$ $<$ 10$^{10}$ M$_{\sun}$). We propose dust stripping and heating processes which could be responsible for the unusually warm characteristic dust temperatures. A normal star-forming galaxy would need 30--50\% of its dust removed (preferentially stripped from the outer reaches, where dust is typically cooler) to recover a SED similar to a ``warm dust" galaxy. These progenitors would not require a higher IR-luminosity or dust mass than the currently observed normal star-forming population.

\end{abstract}

\keywords{Galaxies: clusters: general --  galaxies: star formation -- infrared: galaxies}

\section{Introduction}

High energy photons (i.e. ultraviolet light) heat dust grains and are re-emitted at longer wavelengths, such that far-infrared (FIR) observations are an effective probe of dust properties. In galaxies lacking an IR-bright active galactic nucleus (AGN), the dominant heat source are young stellar populations, and hence FIR flux also traces star formation (SF). The spectral energy distribution (SED) of a dust cloud can be modeled by a composite of many modified blackbodies (greybodies), as the radiative temperature of each grain is a function of size and composition. With coarse photometric data, the dust component is often treated as a single greybody at a fixed characteristic temperature and emissivity, encapsulating the typical dust properties of the galaxy (see discussions in e.g. \citealt{hil83-267,net09-1824}).

Observing in the near and mid-infrared, {\it Spitzer} glimpses the Wien's law (blue-ward) side of the dust component. Generally, MIPS 70 and 160 \micron{} data are too shallow to detect cluster galaxies other than the nearest and brightest (e.g. Shapley Supercluster at z$<$0.05; \citealt{hai11-127}). In contrast, the superb sensitivity of MIPS at 24 \micron{} has been the {\it de rigueur} probe of obscured star formation. Using a small number of local luminous infrared galaxies (LIRGs) as a template library, \citet{rie09-556} derived a simple, yet powerful, formula to extrapolate obscured star formation rate (SFR) directly from 24 \micron{} flux alone.

Cluster member galaxies are typically quiescent early-types with no significant star formation, as depicted by the morphology--density \citep{dre80-351} and SF--density  \citep{dre85-481,pog99-576} relations. Ultraviolet observations do, however, reveal low-level recent star formation in some optical-red-sequence sources \citep{yi05-111,raw08-2097}. Infrared imaging indicates that galaxies (in small groups) accrete onto a cluster along filamentary structures, where gravitational interactions between the galaxies, rather than cluster-potential processes, stimulate star formation \citep{fad08-9,koy08-1758,hai11-145}. So while SF does concentrate in pockets on the periphery of clusters, the star formation rate of star-forming cluster galaxies is no different from identically selected galaxies at lower densities \citep{gea11-177}. Similarly, elevated number counts, but no enhancement in individual SFRs, are reported in the most extreme environments, such as massive cluster mergers \citep{dep07-2209,joh08-289,chu10-1536}. Once star formation in a cluster member ceases, most likely due to gas exhaustion before completing the first in-fall pass \citep[e.g.][]{tre03-53}, further gas cooling is prevented by the hotter intracluster medium of the dense core, giving rise to the observed density relations. An obvious exception are the brightest cluster galaxies at the center of relaxed `cool-core' clusters, which exhibit star formation rates up to $\sim$100 M$_{\sun}$ yr$^{-1}$ fueled by inflowing cool cluster gas \citep{raw12-tmp}.

The {\it {\it Herschel} Space Observatory} \citep{pil10-1} directly constrains the full FIR dust component for a large sample of sources. A small number of early studies analyzed the FIR signatures of cluster galaxies at z$\sim$0.2--0.3. Wide-field analysis recovered a peak in star formation rate at $\sim$$r_{\rm virial}$ ($\sim$2--2.5 Mpc; \citealt{smi10-18}, \citealt{per10-40}). However, the real power of {\it Herschel} is in the constraint of dust characteristics beyond merely IR luminosity. \citet{raw10-14} and \citet{per10-40} reported an unexpected cluster population displaying a FIR SED shape unlike local templates. Specifically the sources exhibited excess flux blue-ward of the dust peak, but were not associated with infrared-bright AGN. \citet{raw10-14}, an analysis of science demonstration phase (SDP) {\it Herschel} observations of the massive, merging Bullet Cluster found that $\sim$40\% of star-forming cluster galaxies have elevated $S_{100}$/$S_{24}$ compared to local templates. Similar sources were not identified in field surveys at comparable redshift, or in high-redshift samples \citep[e.g.][]{elb10-29,rex10-13}.

In the current study, far-infrared observations from the Open Time Key Program, the ``{\it Herschel} Lensing Survey" \citep{ega10-12} are used to explore the dust properties of galaxies in two massive clusters at z$\sim$0.3. The paper has the following structure: Section \ref{sec:obs} introduces the target clusters and data; Section \ref{sec:analysis} describes photometric analysis and SED fitting; Section \ref{sec:results} explores the IR luminosity, dust properties and physical interpretation; Section \ref{sec:summary} summarizes the primary conclusions. An in-depth examination of the SED fits is detailed in the Appendices.

\section{Observations}
\label{sec:obs}

\placefigure{fig:footprints}

\begin{figure*}
\includegraphics[angle=270,scale=0.55]{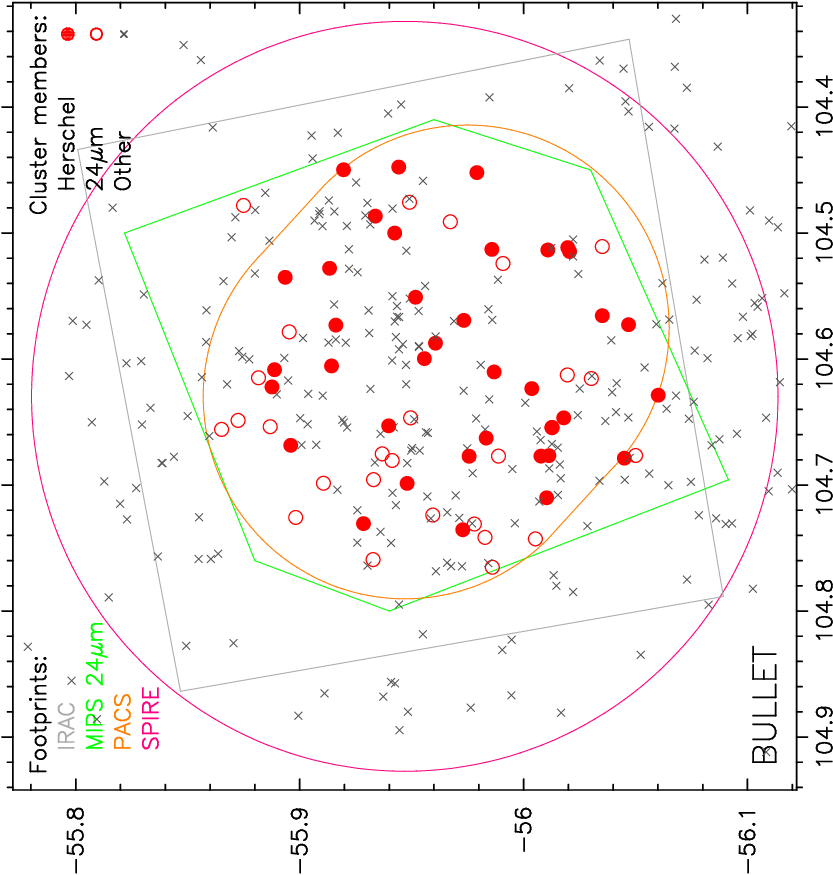}
\includegraphics[angle=270,scale=0.55]{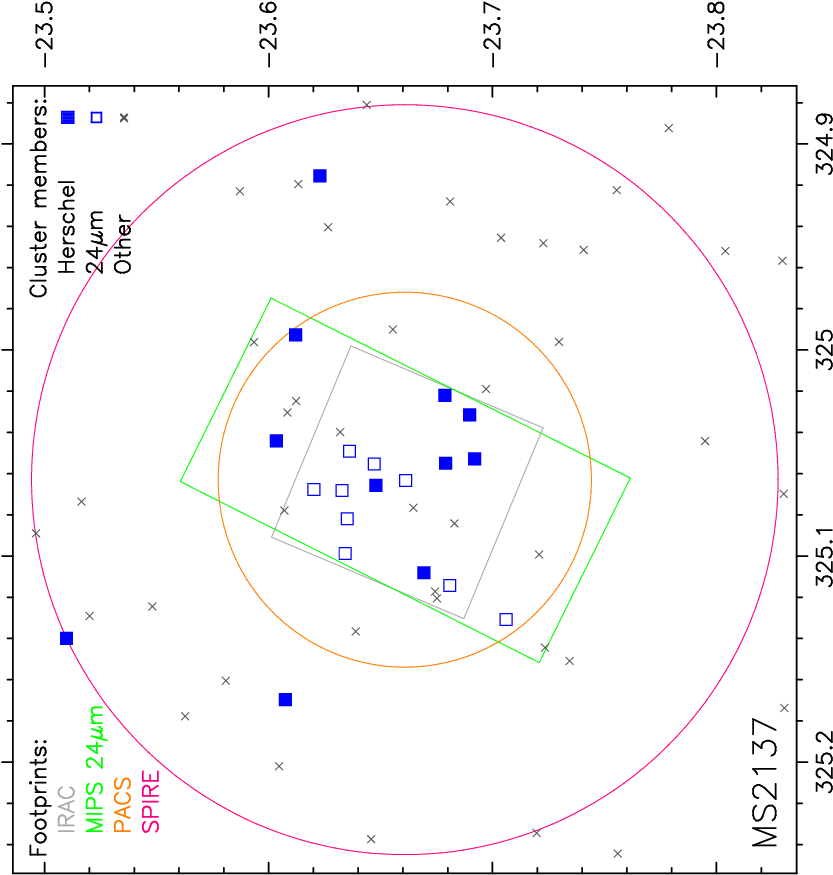}
\caption{20'$\times$20' fields aligned to the nominal cluster centers: Bullet Cluster ({\it left}) and MS2137 ({\it right}). Imaging footprints are shown as outlines: IRAC$=$grey, MIPS 24 \micron{}$=$green, PACS$=$orange, SPIRE$=$magenta. Spectroscopically confirmed cluster members are indicated: crosses are undetected at $\lambda$ $\ga$ 24 \micron{}; unfilled symbols are detected at MIPS 24 \micron{} but not by {\it Herschel}; filled symbols are detected by {\it Herschel}. Initiating the convention for plots in this paper, Bullet Cluster$=$red circles, MS2137$=$blue squares. The larger number of Bullet Cluster detections results from more substantial spectroscopy.}
\label{fig:footprints}
\end{figure*}

The ``{\it Herschel} Lensing Survey" (HLS; \citealt{ega10-12}) is designed to exploit the lensing effect of massive clusters, pushing beyond the nominal confusion limit to observe intrinsically faint, high-redshift sources, e.g. \citet{rex10-13}, \citet{com12-tmp}. To this end, HLS comprises deep PACS \citep{pog10-2} and SPIRE \citep{gri10-3} imaging (100--500 \micron) of 44 massive cluster cores, 0.2 $\la$ $z$ $\la$ 0.5.

In this analysis, we concentrate on galaxies within two massive systems: the Bullet Cluster and MS2137.3--2353. These clusters have substantial multi-wavelength photometric coverage, alongside 100s of spectroscopic redshift measurements. Although photometric redshifts derived from optical--FIR photometry are sufficiently accurate to identify e.g. the high redshift population \citep{per10-15}, isolation of cluster members requires precise spectroscopic redshifts. Furthermore, redshift and dust temperature are degenerate when examining the FIR SED. By placing a 30 K greybody at $z$$\sim$0.3, and then assuming various redshifts within the typical photometric redshift error ($\Delta z$$\sim$0.1; \citeauthor{per10-15}), we estimate the associated temperature uncertainty to be $\ga$10 K.

In this section we detail relevant observations, while Figure \ref{fig:footprints} displays the spatial coverage of the infrared imaging.

\subsection{The Bullet Cluster (1E 0657-56)}
\label{sec:bullet}

\citet{raw10-14} presented a preliminary analysis of the HLS SDP data available for the Bullet Cluster (RA $=$ 06:58:31.10, Dec $=$ --55:56:49.0; $z$ $=$ 0.296). The system is a recent collision of two clusters \citep{mar02-27}, perpendicular to the line of sight \citep{mar04-819}, offering a unique laboratory for the study of SF within a dense dynamic environment. X-ray emission shows a supersonic bow shock preceding the hot gas, while the weak lensing mass profile indicates that this X-ray bright component lags behind the sub-cluster galaxies due to ram pressure \citep{clo04-596,clo06-109}. Ram pressure from the merger event does not appear to impact the star formation rate of galaxies in the vicinity of the hot X-ray gas \citep{chu09-963,raw10-14}.

As described in the introduction, early {\it Herschel} cluster observations identified a surprising population of sources with an apparent excess at 100 \micron{}. We obtained additional PACS coverage of the Bullet Cluster to explore these sources further.

\subsubsection{{\it Herschel}/PACS observations}

The SDP PACS observations (presented in \citealt{ega10-12,raw10-14}) consisted of two orthogonal scan maps, each comprising 18 repetitions of thirteen parallel 4' scan legs. Total observing time was 4.4 hours, with 1500 sec/pixel integration time. PACS images are produced using a modified version of the standard reduction package, HIPE v6.0 \citep{ott10-139}. High-pass filtering of the time-stream is necessary to remove the 1/f noise drift of the PACS bolometers, and we implement a simple masking algorithm to avoid ringing of bright sources. Raw PACS data includes a large number of usable frames beyond the science scan legs (turnaround data). We include all frames with speeds within 5\% of the nominal scan speed. As the detector footprint is comparable in size to the final map, the additional frames increase signal-to-noise in the central regions, as well as overall spatial extent. Thus, SDP coverage of the central 8'$\times$8' reached a 3$\,\sigma$ depth of 3.1 and 5.8 mJy in the 100 and 160 \micron{} bands.

The new, augmented PACS imaging is an extension on the SDP data in three senses: (1) the 100 and 160 \micron{} footprints are approximately doubled ($\sim$8'$\times$15') by a new observation (identical configuration), to the south-west of the original pointing, chosen for the number of potentially interesting sources identified in the larger SPIRE images; (2) both pointings were observed in a single, deep 70 \micron{} observation, probing the SED between existing 24 and 100 \micron{} data; (3) the 70 \micron{} pointing simultaneously observes 160 \micron{}, providing extra depth in the longest PACS band. The 70 \micron{} map consists of two orthogonal scan maps of 21 repetitions each. The first has 28 parallel scan legs of 5.2' length, while the second has sixteen parallel 9' scan legs, with the pointing centers chosen to create a rectangular map ($\sim$15'$\times$8' with turnaround data). Total observing time is 10.5 hours, with $\sim$2100 sec/pixel integration time. The final 70 \micron{} map is actually larger than the two 100 \micron{} footprints, resulting in a significant number of sources near the edge of the PACS coverage with 70 and 160 \micron{} photometry only.

The three final maps reach mean 3$\,\sigma$ detection limits of 2.7, 2.8, 3.8 mJy at 70, 100, 160 \micron{} respectively (beam sizes of 5.2, 7.7, 12"; detection limits calculated using the method described in \citealt{per10-15}). These maps are all above the (3$\,\sigma$) confusion limits of 0.07, 0.3, 2.3 mJy \citep{ber10-30}.

\subsubsection{{\it Herschel}/SPIRE observations}
\label{sec:bullet_spire}

The Bullet Cluster has not been re-observed by SPIRE since SDP, which used large map mode: 20 repetitions each with a nominal length and height of 4'. The total observing time was 1.7 hours, with 17 sec/pixel. The images were produced via the standard reduction package HIPE v5.0, but include all turnaround data with speeds $\ga$0.5"/s to increase signal-to-noise in the outer regions. Final maps are complete to a cluster-centric radius $>$9', with 3$\,\sigma$ detection limits of 8.2, 9.6, 12.4 mJy at 250, 350, 500 \micron{} respectively \citep{per10-15}. All three bands are confusion limited (3$\,\sigma_{\rm conf}$ $\approx$ 17, 19, 20 mJy; \citealt{ngu10-5}), with beam sizes of 18, 25, 36".

\subsubsection{Additional imaging}

As a consequence of the unique cluster characteristics, and a well known, lensed background source \citep{gon09-525}, there is a wealth of available ancillary data for the Bullet Cluster. The central 13'$\times$13' (including the entire PACS field) is covered by 5-band {\it Spitzer} imaging \citep{gon09-525} from IRAC (3.6, 4.5, 5.8, 8.0 \micron{}) and MIPS (24 \micron{}), with 3$\,\sigma$ detection limits of 2, 3, 5, 8 and 50 $\mu$Jy respectively. Also available are deep wide-field images in $BVR$ from Magellan/IMACS \citep{clo06-109}, $YJH$ from VLT/HAWK-I (PI: J.-G. Cuby) and X-ray from a 0.5 Msec {\it Chandra} integration \citep{mar06-723}. See the appendix of \citealt{ega10-12} for further details.

\subsubsection{Spectroscopic coverage}

Spectroscopic redshifts in the Bullet Cluster field originate from three campaigns: Magellan IMACS multi-slit (856 spectra; \citealt{chu10-1536}), CTIO Hydra multi-fiber (202; D. Fadda et al. in preparation) and VLT FORS multi-slit (14; J. Richard, private communication). The merged catalog comprises 929 sources within the SPIRE field, including 371 cluster members (0.28 $<$ z $<$ 0.31). 47 cluster galaxies are included in multiple spectroscopic datasets, with no membership disagreement. Figure \ref{fig:footprints} indicates the location of spectroscopic cluster members with respect to the infrared imaging coverage.

\subsection{MS2137.3--2353}

MS2137.3--2353 (RA $=$ 21:40:15.10, Dec $=$ --23:39:39.0; $z$ $=$ 0.313; hereafter MS2137) is an ideal counterpoint to the Bullet Cluster. Located at similar redshift, the massive cluster is an approximately spherical system \citep{san02-129}, with an isolated BCG coincident with the X-ray center (a reliable indicator of an undisturbed cluster; e.g. \citealt{san09-1698}).

\subsubsection{{\it Herschel} observations}

MS2137 has HLS FIR coverage comparable to the Bullet Cluster SDP, with 8'$\times$8' PACS maps at 100 and 160 \micron{} and three-band SPIRE coverage (250--500 \micron{}). The PACS sensitivities (3$\,\sigma$ limits of 2.5 and 4.6 mJy) are marginally lower than in the Bullet Cluster SDP maps as the efficiency of PACS improved during the first year of operation. The nominal detection limits of the SPIRE maps are consistent with the Bullet Cluster (8.0, 9.3, 10.3 mJy). All {\it Herschel} instrument configurations and data reduction are as described for the Bullet Cluster in Section \ref{sec:bullet}.

\subsubsection{Additional imaging}

The spatial coverage in the {\it Spitzer} bands is more restricted for MS2137 than the Bullet Cluster, as shown by Figure \ref{fig:footprints}. However, {\it Spitzer} observations did occur during the cryogenic phase (PIs: G. Rieke, E. Egami), so all four IRAC channels and MIPS 24 \micron{} are available, with the same sensitivities as in the Bullet Cluster. The entire SPIRE field is enclosed by deep $BVRIZ$ Subaru Suprime-Cam observations (PI: D. J. Sand).

\subsubsection{Spectroscopic coverage}

MS2137 was targeted in two complementary spectroscopic campaigns using the Magellan Telescopes: 115 24 \micron{}-selected galaxies in the cluster core, observed using LDSS3 (PI: E. Egami); IMACS masks targeting {\it Herschel} sources (PI: T. Rawle), exploiting the larger field of view to observe SPIRE detections beyond the LDSS3 field. The merged catalog contains 231 sources within the SPIRE footprint, of which 78 are in the cluster (0.30 $<$ z $<$ 0.33). Nine sources were observed by both instruments, with no disagreement.

\subsection{Field samples}

We compile low-density comparison samples with the same observed photometry as the cluster population. These field samples comprise all galaxies spectroscopically confirmed to be in the foreground of either cluster (0.05 $<$ z $<$ 0.26). The most local galaxies (z $<$ 0.05) are excluded as our filtered PACS maps are not optimized for extended sources. Background sources are also not included to avoid the additional uncertainty of lensing on the intrinsic luminosity. Although we mostly employ these two foreground samples simultaneously, we maintain the distinction between the fields because of the difference in photometric coverage (specifically a lack of 70 \micron{} for MS2137). The field and cluster sources are treated identically at every stage of analysis.

We also compare to large blank-field samples from \citet[][BBC03]{bla03-733} and \citet[][H10]{hwa10-75}, selecting the 129 sources in the redshift range 0.05 $<$ z $<$ 0.5. BBC03 is a compilation of earlier studies using mostly IRAS data in the infrared. H10 comprises two subsamples, an SDSS-selected sample observed by AKARI (predominantly z $<$ 0.1), and a {\it Herschel}-detected sample in the GOODS-North field (mostly higher redshift). Both contribute sources in the required redshift range for this comparison. See Sections \ref{sec:fitting} and \ref{sec:lt} for more details on the suitability and usage of the external field samples.

\section{Analysis}
\label{sec:analysis}

\begin{table}
\caption{Final sample sizes\label{tab:samples}}
\begin{tabular}{lccc}
\tableline
 & \multicolumn{1}{c}{Total\tablenotemark{a}} & \multicolumn{1}{c}{{\it Herschel}\tablenotemark{b}} & \multicolumn{1}{c}{Low $L_{\rm IR}$\tablenotemark{c}}\\
\tableline
Bullet Cluster & 64 & 37 & 37 \\
MS2137 & 20\tablenotemark{d} & 11 & 9 \\
Bullet field & 34 & 20 & 19 \\
MS2137 field & 15 & 11 & 9 \\
\end{tabular}
\tablenotetext{a}{Spectroscopically-confirmed cluster members with 24 \micron{}\\ detection; see footnote \textit{d}}
\tablenotetext{b}{...and at least two {\it Herschel} detections for FIR SED fitting}
\tablenotetext{c}{...and $L_{\rm IR}$ $<$ 2$\times$10$^{11}$ L$_{\odot}$ (for Table \ref{tab:nwarm})}
\tablenotetext{d}{Includes three isolated spectroscopic cluster members\\ detected by SPIRE beyond the MIPS 24 \micron{} coverage\\ (see Section 3.1)}
\end{table}

\subsection{Multi-wavelength counterparts}
\label{sec:counterparts}

The final samples are derived by matching MIPS 24 \micron{} and (optical) spectroscopic catalogs.  A 24 \micron{} detection (beam size 6") could be the counterpart to any optical imaging source within the MIPS rms pointing error (1.4"), and all were initially viewed as potential matches. The deep IRAC maps contain many more sources than MIPS 24 \micron{}, including a significant fraction of optical detections. The most likely optical counterpart to any MIPS source can be selected by interactive examination of flux and position through all available intermediate (i.e. near-infrared and IRAC) bands. The optical imaging counterparts are then cross-referenced to the spectroscopic catalogs to derive the matched MIPS--spectroscopic catalog. A conservative approach is adopted to ensure MIPS sources are not assigned incorrect redshifts; spectroscopically-confirmed optical sources with unsure matches in the IRAC or MIPS bands are not included in the final samples.

Given the detection limits of infrared imaging available for the two clusters, typical SEDs show that any source at z$\sim$0.3 with a significant {\it Herschel} detection, will also be detected at 24um. Identifying the corresponding sources in the un-confused PACS bands (beam size $\sim$8") is straightforward, and the small number of targets in this study allows for individual examination of every source. An automated or statistical counterpart algorithm may miss sources with unusual SED shapes. In the confusion-limited SPIRE bands, counterparts are not located directly, as flux is assigned to all MIPS sources simultaneously (described further in the following section). Three bright {\it Herschel} sources in MS2137, which lie beyond the MIPS 24 \micron{} coverage are matched unambiguously to isolated optical sources identified in Subaru imaging. These sources are included despite the lack of 24 \micron{} photometry; removing them from the sample does not affect the results.

The final sample sizes are presented in Table \ref{tab:samples}, with the cluster members also plotted in Figure \ref{fig:footprints}. The Bullet Cluster analysis presented in \citet{raw10-14} comprised 23 sources, 14 of which had only SPIRE photometry. The new PACS imaging completes 70--160 \micron{} coverage for all these sources, while also increasing the FIR-detected cluster sample by a further $>$50\%.

\subsection{Photometry}
\label{sec:phot}

For {\it Spitzer} and PACS, where blending and source confusion is minimal, fluxes are measured using simple aperture photometry via the {\sc Sextractor} package \citep{ber96-393}. Aperture corrections are applied to each band, based on instrument handbook values. As recommended for PACS imaging produced by the pipeline (and calibration product) used in this study\footnote{PACS observers' manual; {\it http://herschel.esac.esa.int/Docs/PACS/html/pacs\_om.html}}, photometry is calculated in 12", 12", 22" radius apertures (at 70, 100, 160 \micron{} respectively) with aperture corrections of 0.886, 0.886, 0.916. Background subtraction is not required prior to aperture photometry in PACS maps, as the high-pass filtering artificially sets the mean background to zero. Flux uncertainties are estimated by {\sc Sextractor} from the global sky rms error.

In the SPIRE bands, where background source confusion is above the instrument noise, PSF fitting is the favored technique for measuring fluxes. We employ the {\sc Iraf} routine {\sc Allstar}, simultaneously fitting a model PSF to all MIPS 24 \micron{} positions. The use of a deep prior catalog removes a large fraction of the contaminating background, minimizing the need to de-boost SPIRE fluxes. Significant residual flux (e.g. from faint high-z sources below the 24 \micron{} detection limit) would be identified by unphysical discontinuities in the final SED. In practice, for PACS sources at z$<$0.5 the detection rate in the SPIRE 250 \micron{} band is $\sim$50\%. At 350 and 500 \micron{} the Rayleigh--Jeans tail becomes very faint, and the detection rate is even lower at 20\% and 5\% (3 sources) respectively.

\subsection{Infrared SEDs}
\label{sec:fits}

\subsubsection{SED fitting}
\label{sec:fitting}

Infrared luminosity is derived via the best-fitting template to all MIPS and {\it Herschel} photometry (i.e. $\lambda_{\rm obs}$ $\ge$ 24 \micron{}). IRAC data is not included in the fits as these wavelengths are dominated by poorly constrained PAH features, and only sources with at least two {\it Herschel} detections are analyzed. We use the \citet[R09]{rie09-556} and \citet[CE01]{cha01-562} templates, derived from a small number of star-formation dominated local (U)LIRGs interpolated to produce a library of galaxy spectra classified by IR luminosity. Template shape varies significantly with luminosity, in the sense that higher luminosity dust peaks at shorter wavelengths. As we are interested in studying this shape for cluster galaxies, we ignore the nominal luminosity class of the templates to find the best fit via a least squares minimization (from the {\sc Python Scipy.Optimize} package), and allow a free-floating normalization factor to scale the template to the intensity of the source. Hereafter the luminosity class of the best-fitting template will be referred to as the ``template luminosity", $L_{\rm template}$.

The true total infrared luminosity of each galaxy ($L_{\rm IR}$) is derived by integrating the scaled, best-fitting template over the (rest frame) wavelength range $\lambda_{\rm 0}$ $=$ 8--1000 \micron. Uncertainty in the luminosity is estimated by Monte Carlo simulations of the fitting procedure. The star formation rate (SFR$_{\rm IR}$) is calculated from $L_{\rm IR}$ following the simple relation presented in \citet{ken98-189}\footnote{SFR$_{\rm IR}$ [$M_{\sun}$ yr$^{-1}$] $=$ 4.5$\times$10$^{-44}$ $L_{\rm IR}$ [erg s$^{-1}$]}.

The characteristic temperature of the dust component is calculated by fitting a single-temperature greybody of the form
\begin{equation}
S_{\nu} = N{\nu}^{\beta}B_{\nu}(T_{\rm dust})
\end{equation}
where $S_{\nu}$ is flux density, $N$ is a free-floating normalization factor, $\beta$ is dust emissivity index and $B_{\nu}(T_{\rm dust})$ is the Planck blackbody radiation function for a source at temperature $T_{\rm dust}$. Typically, $\beta$ is observed in the range $\sim$1--2 (e.g. $\beta$ $=$ 1.8 in the solar neighborhood, \citealt{pla11-19}). We fix $\beta$ $=$ 1.5 to mirror the analysis of our external comparison samples BBC03+H10\footnote{BBC03 and H10 employ exactly the same greybody formulation as we use, with $\beta$=1.5}, although we note that $\beta$ $=$ 2 is another popular choice in the literature. $\beta$$T_{\rm dust}$ remains approximately constant, so using $\beta$=2.0 would systematically reduce the derived temperatures by 10\%. $T_{\rm dust}$ for a z$\sim$0.3 cluster galaxy is usually well constrained, as the SED peaks between 100 and 160 \micron{} for dust at $\sim$30 K. In contrast to the template fits described above, 24 \micron{} photometry is not included in the greybody fit. Sources with fewer than two {\it Herschel} detections are excluded from the greybody fit procedure.

\subsubsection{Robustness of SED fits}
\label{sec:valid}

First we compare the derived infrared luminosity from the best fit CE01 and R09 templates. The mean ratio $L_{\rm IR}$(CE01)/$L_{\rm IR}$(R09) $=$ 0.90 $\pm$ 0.02, with an rms scatter of 0.08 dex. $L_{\rm IR}$ is generally independent of template set, with a marginal excess in the mid-infrared of the best fit CE01 template for sub-LIRG galaxies. Appendix \ref{app:24um} expands this analysis by comparing $L_{\rm IR}$(R09) to values extrapolated from 24 \micron{} data alone.

We now examine the dependence of our fits (and hence the derived values of the primary parameters $L_{\rm IR}$ and $T_{\rm dust}$) on the availability of data at various wavelengths. We test the effect of non-detection in SPIRE bands by temporarily excluding photometry at $\lambda$ $\ge$ 250 \micron{}. $L_{\rm IR}$ and $T_{\rm dust}$ typically differ by $<$0.1 dex and $<$1 K (respectively) despite the poorer constraint on the Rayleigh--Jeans tail. The difference is larger ($\sim$0.3--0.4 dex and 3 K) for sources with particularly cool characteristic dust temperatures ($T_{\rm dust}$ $<$ 25 K), where the dust component peak lies closer to the SPIRE bands. Typically, the dust SED maximum lies within the PACS bands for sources at the cluster redshift and below, and therefore PACS data is the primary constraint on the best fit.

\begin{figure}
\includegraphics[angle=0,scale=0.64]{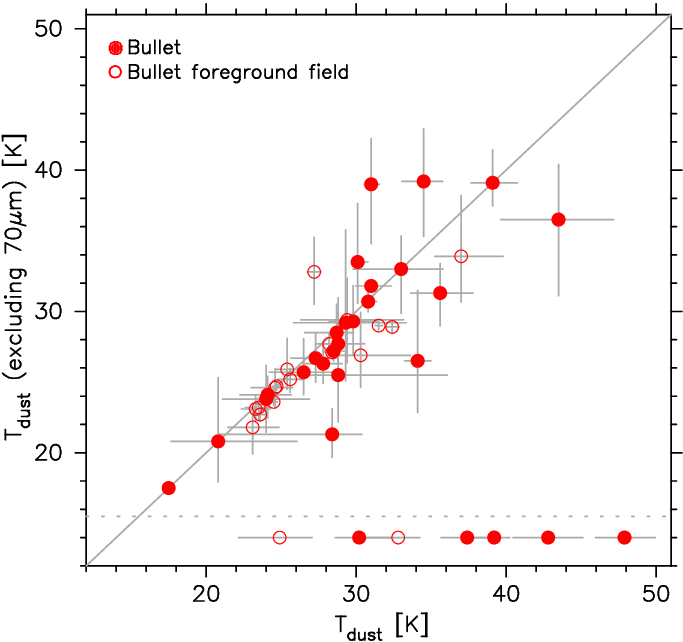}
\caption{The effect of 70 \micron{} data on the characteristic dust temperature $T_{\rm dust}$ for sources in the Bullet Cluster (filled red circles) and foreground field (open red circles). There is significantly more scatter for warmer sources, where 70 \micron{} is required to constrain the peak. When 70 \micron{} photometry is excluded, sources below the dotted line no longer have sufficient data for a greybody fit, and are plotted at 14 K for convenience.}
\label{fig:70um}
\end{figure}

A major difference between the Bullet Cluster and MS2137 data is the lack of 70 \micron{} coverage in the latter. For sources in the Bullet Cluster (and foreground field), we compare parameters derived from fits excluding 70 \micron{} photometry, to those derived in Section \ref{sec:fits}. $L_{\rm IR}$(no 70 \micron{})/$L_{\rm IR}$ = 1.0, with an rms scatter of 0.08 dex (for either template set). Figure \ref{fig:70um} compares $T_{\rm dust}$ derived when including/excluding 70 \micron{} photometry, and shows an increased scatter for sources with $T_{\rm dust}$ $\ga$ 30 K. Although $S_{\rm 160}$/$S_{\rm 100}$ is a reasonable tracer of $T_{\rm dust}$ (see Appendix \ref{app:lum} for more details, especially Figure \ref{fig:pacscols}), above $\sim$30 K, 70 \micron{} photometry adds significant additional information about the SED shape. Excluding sources $T_{\rm dust}$ $>$ 30 K from the previous $L_{\rm IR}$ test reduces the rms scatter to 0.04 dex (from 0.08 dex). These results suggest that 70 \micron{} data is required to fully constrain the peak of dust component in the warmest cases; without 70 \micron{} photometry, the uncertainty on $L_{\rm IR}$ and $T_{\rm dust}$ in warm dust components is approximately doubled. We return to this bias in Section \ref{sec:sig}.

\subsubsection{Observed SEDs}

\begin{figure*}
\includegraphics[angle=270,scale=0.67]{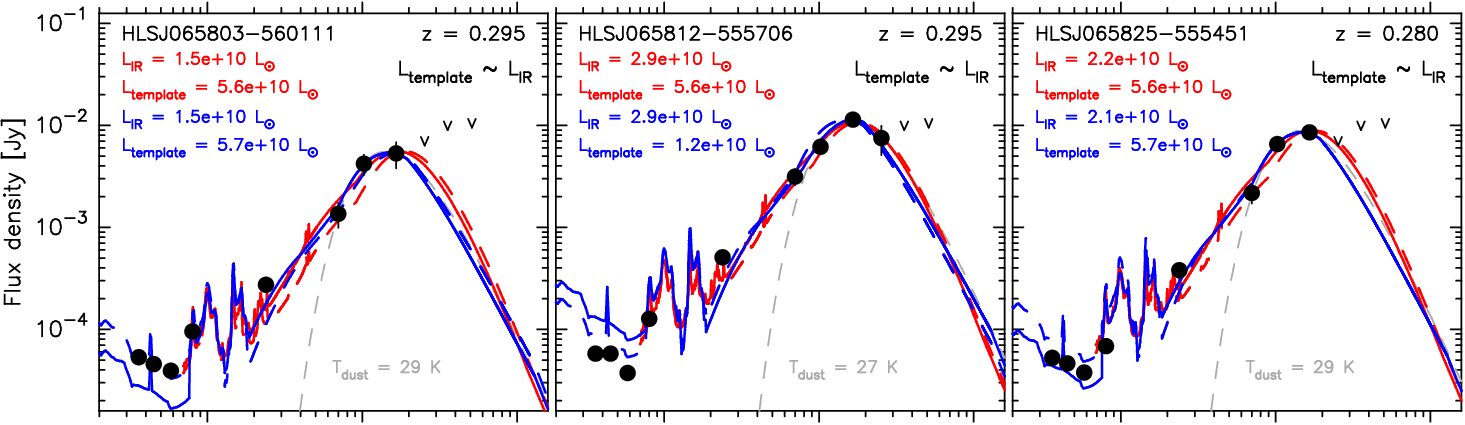} \\
\\
\includegraphics[angle=270,scale=0.67]{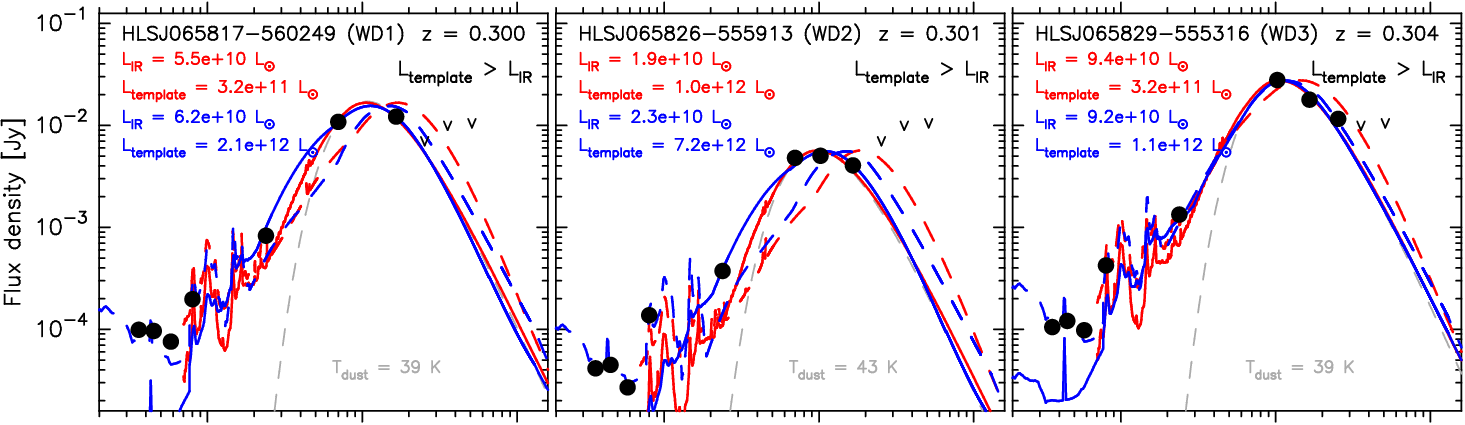} \\
\includegraphics[angle=270,scale=0.67]{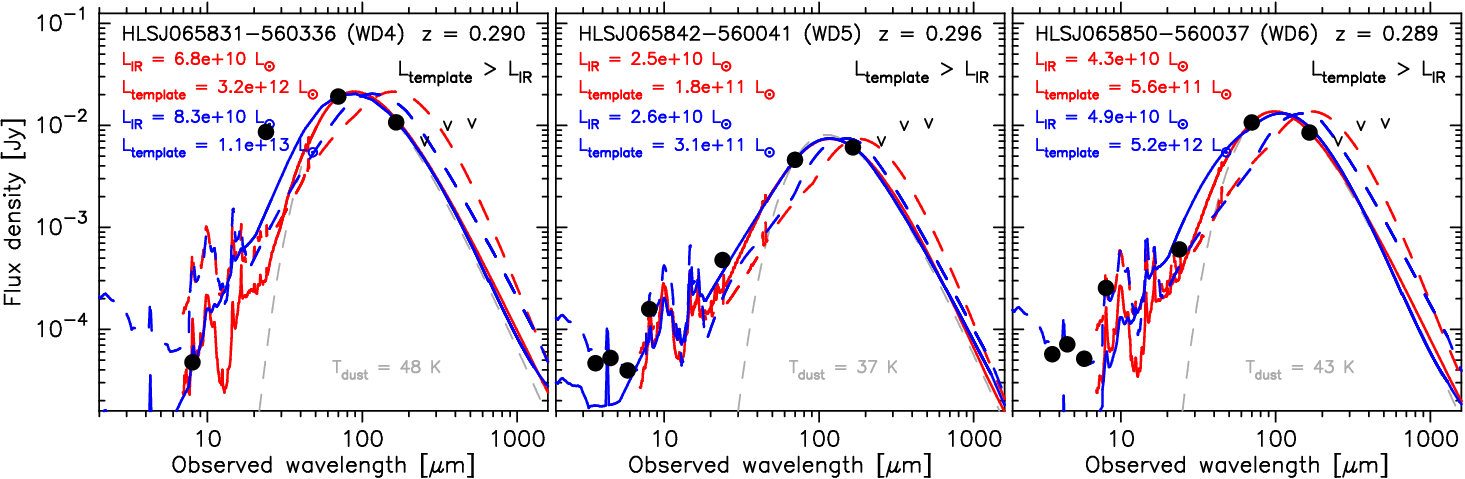}
\caption{IR SEDs of spectroscopically-confirmed member galaxies in the Bullet Cluster. {\it Uppermost row:} Three examples of normal star-forming galaxies, for which $L_{\rm template}$ $\sim$ $L_{\rm IR}$; {\it Lower rows:} The six `warm dust' galaxies with $L_{\rm template}$ $>$ $L_{\rm IR}$. In each panel, observed photometry is plotted in black. Best fitting templates are shown as solid lines (red$=$R09, blue$=$CE01), with $L_{\rm IR}$ and $L_{\rm template}$ given in matching colors in the upper-left. The template with $L_{\rm template}$$=$$L_{\rm IR}$ is added as a dashed line normalized to the same peak flux as the solid line. The best-fitting greybody is displayed as a grey dashed line, with the characteristic dust temperature shown to the lower-right.}
\label{fig:seds}
\end{figure*}

\begin{figure*}
\centering
\includegraphics[angle=0,scale=0.9]{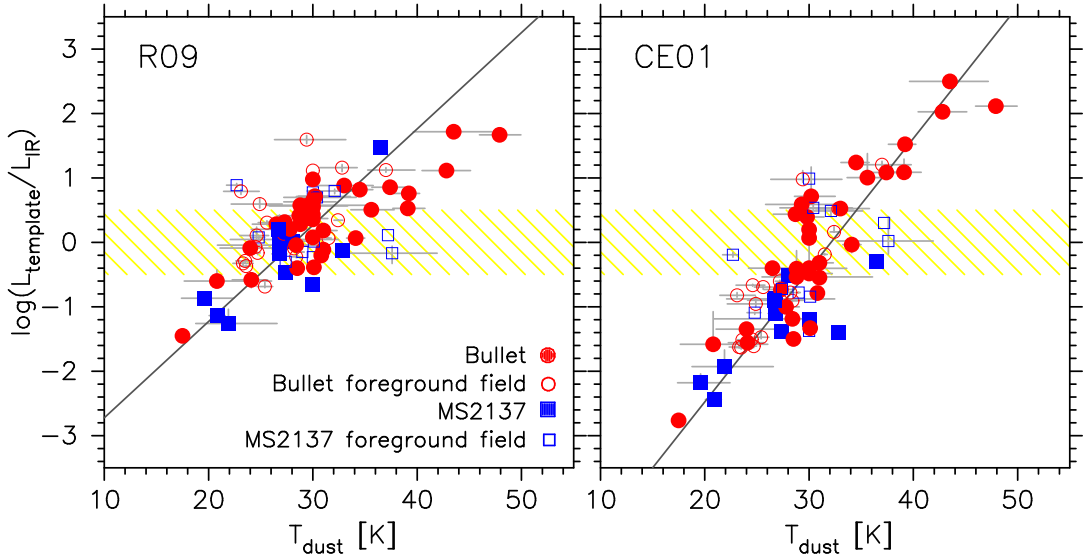}
\caption{The template mismatch, log($L_{\rm template}$/$L_{\rm IR}$), compared to characteristic dust temperature ($T_{\rm dust}$) derived from the best-fitting greybody (left panel$=$R09; right panel$=$CE01). Bullet Cluster$=$red filled circles, MS2137$=$blue filled squares. Field samples in corresponding open symbols. Solid lines show the best linear fit to the data, implying a strong correlation between the parameters. The (yellow hatched) shaded regions guide the eye to sources for which $\vert {\rm log(}L_{\rm template}/L_{\rm IR}{\rm )}\vert$ $<$ 0.5 (i.e. those without a significant template mismatch).}
\label{fig:dusttemp}
\end{figure*}

Figure \ref{fig:seds} displays IR SEDs for a selection of sources in the Bullet Cluster. In all cases the best fitting template is a good description of the overall shape of the observed SED. The galaxies shown in the upper row are best fit by templates with a luminosity class consistent with the integrated luminosity (i.e. $L_{\rm template}$$\sim$$L_{\rm IR}$). In other words, the shape of the dust component is the same as a local galaxy of similar luminosity. Although the best fitting template for the SEDs in the lower two rows generally correspond well with the observations (the lower-left panel; marked WD4is an exception), $L_{\rm template}$ is more than an order of magnitude higher than the integrated $L_{\rm IR}$. The shape of the dust component resembles a much more luminous local template. Figure \ref{fig:dusttemp} displays the template mismatch for all the {\it Herschel} sources. \citet{raw10-14} attempted to explain the discrepancy via an excess in S$_{100}$/S$_{24}$. Appendix \ref{app:lum} presents a detailed exploration of this discrepancy. We suggest that the peak wavelength (traced by e.g. S$_{160}$/S$_{100}$) is a much more successful probe of the unusual behavior. Furthermore, Figure \ref{fig:dusttemp} shows that for both template sets, the discrepancy [i.e. log($L_{\rm template}$/$L_{\rm IR}$) $\ne$ 0] is well correlated with characteristic dust temperature, a parameter the templates effectively keep constant at a given luminosity. Sources with the largest template discrepancy have unusually warm or cold characteristic dust temperatures for their luminosity.

\section{Results and discussion}
\label{sec:results}

Throughout the remainder of this study, we use the infrared luminosity derived from R09 templates. Our conclusions would not change if we instead adopted the CE01 templates, as noted in Section \ref{sec:valid}.

\subsection{IR luminosity -- temperature relation}
\label{sec:lt}

\begin{figure*}
\centering
\includegraphics[angle=0,scale=0.84]{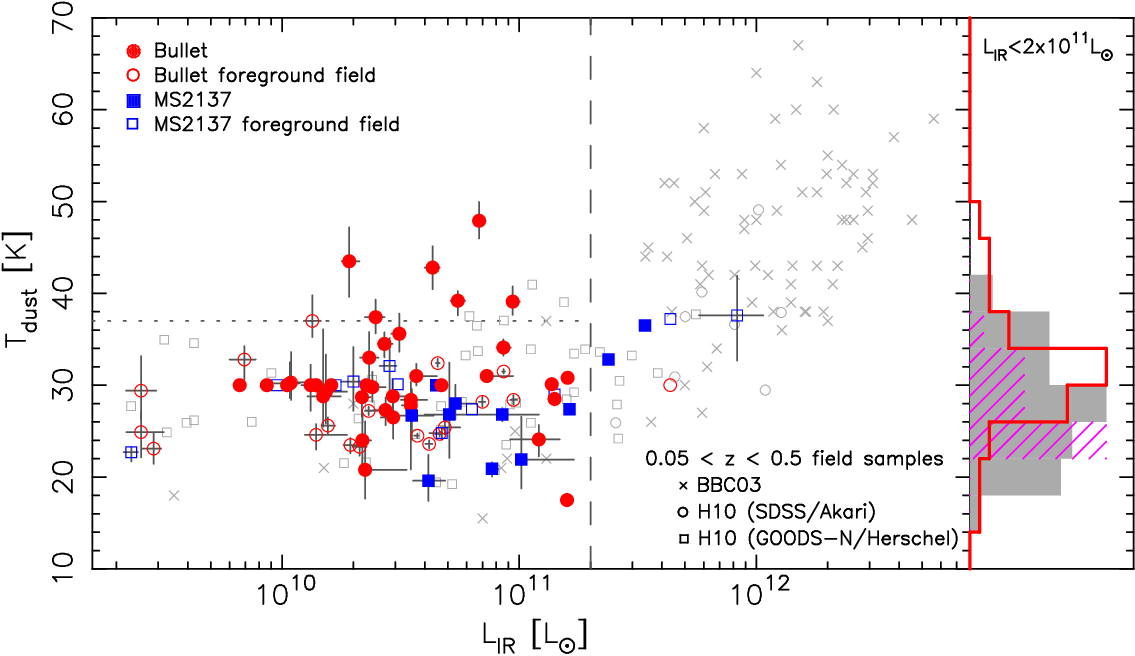}
\caption{{\it Main panel:} IR luminosity -- dust temperature relation for the cluster (red filled circles$=$Bullet, blue filled squares$=$MS2137) and field samples (corresponding open symbols). $L_{\rm IR}$ is derived from the R09 templates, and $T_{\rm dust}$ from a single greybody. Grey symbols display 0.05 $<$ z $<$ 0.5 sources from \citet[][BBC03]{bla03-733} and \citet[][H10]{hwa10-75}. The vertical dashed line guides the eye to the dichotomy at $L_{\rm IR}$ $\sim$ 2$\times$10$^{11}$ L$_{\odot}$. Several galaxies in the Bullet Cluster have significantly warmer dust ($T_{\rm dust}$ $>$ 37 K; above the horizontal dotted line) than similar luminosity field galaxies. All but one field galaxies with $L_{\rm IR}$ $<$ 2$\times$10$^{11}$ L$_{\odot}$ have $T_{\rm dust}$ $<$ 33 K. {\it Right panel:} Normalized distribution of dust temperatures for $L_{\rm IR}$ $<$ 2$\times$10$^{11}$ L$_{\odot}$ galaxies in the Bullet Cluster (red outline), Bullet foreground sample (magenta hatching) and BBC03+H10 field (grey solid).}
\label{fig:lir_temp_blain}
\end{figure*}

A relationship between dust temperature and infrared luminosity is often assumed, usually implicitly through the use of templates (e.g. \citealt{dal02-159}, and those in this study). \citet[][BBC03]{bla03-733} presented the observed luminosity--temperature relation, concluding that the scatter was too large to estimate temperature uniquely from luminosity (or {\it vice versa}). BBC03 also highlighted a stark dichotomy in this parameter space. Whilest the brightest LIRGs ($L_{\rm IR}$ $\ga$ 2$\times$10$^{11}$ L$_{\odot}$) at all redshifts were observed to have $T_{\rm dust}$ in the range 30--70 K, less luminous objects rarely have $T_{\rm dust}$ $>$ 40 K with a mean $\sim$30 K. Consequently, the sub-LIRG population should be treated separately, and we adopt a cut at the turning point, $L_{\rm IR}$ $=$ 2$\times$10$^{11}$ L$_{\odot}$.

We explore the distribution of our {\it Herschel} sources in the $L_{\rm IR}$--$T_{\rm dust}$ plane in Figure \ref{fig:lir_temp_blain}. Our field samples (open colored circles and squares) lie within the locus of 0.05 $<$ z $<$ 0.5 sources from BBC03 and H10 (small grey symbols). Two MS2137 field galaxies have $T_{\rm dust}$$\sim$37 K, but also $L_{\rm IR}$ $>$ 4$\times$10$^{12}$ L$_{\odot}$, so are not unusually warm for their luminosity. One possible outlier is the Bullet field source with $T_{\rm dust}$ $=$ 37 K and $L_{\rm IR}$ $\sim$ 1.2$\times$10$^{10}$ L$_{\odot}$ (HLSJ065807--555541). This galaxy falls within a minor foreground over-density at z $=$ 0.234, so may be a group galaxy rather than an isolated field source. All other field galaxies in our samples have $T_{\rm dust}$ $<$ 33 K.

MS2137 cluster members (filled blue squares in Figure \ref{fig:lir_temp_blain}) also lie within the BBC03+H10 distribution. The two warmest sources in MS2137 ($T_{\rm dust}$ $=$ 33--37 K) are the most luminous cluster galaxies in this study, $L_{\rm IR}$ $=$ 2--4$\times$10$^{11}$ L$_{\odot}$. Both are on the periphery of the cluster and neither are covered by PACS, so the dust peak is not well constrained ($T_{\rm dust}$ is likely to be a lower limit). No sub-LIRG sources in MS2137 have $T_{\rm dust}$ $>$ 33 K.

In contrast, the Bullet Cluster (filled red circles in Figure \ref{fig:lir_temp_blain}) contains several obvious outliers in the luminosity--temperature plane, visible as a high-temperature tail in the normalized histogram of sources with $L_{\rm IR}$ $=$ 2$\times$10$^{11}$ L$_{\odot}$, displayed in the right-hand panel of Figure \ref{fig:lir_temp_blain}. A similar tail is not identified in the corresponding field distributions.

\placetable{tab:nwarm}
\begin{deluxetable}{lccccccc}
\tablecolumns{10}
\tablewidth{0pc}
\tablecaption{Significance of the number of warm dust galaxies}
\tablehead{\colhead{$L_{\rm IR}$ $<$ 2$\times$10$^{11}$ L$_{\odot}$} & & \multicolumn{2}{c}{$T_{\rm dust}$$>$40 K} & \multicolumn{2}{c}{$T_{\rm dust}$$>$37 K} & \multicolumn{2}{c}{$T_{\rm dust}$$>$33 K}\\
\colhead{low luminosity subsamples} & \colhead{$n_{\rm sample}$} & \colhead{$n$} & \colhead{$P$} & \colhead{$n$} & \colhead{$P$} & \colhead{$n$} & \colhead{$P$}}
\startdata
Bullet Cluster & 37 & 3 & -- & 6 & -- & 10 & -- \\
MS2137 & 9 & 0 & 38\% & 0 & 13\% & 0 & 3.8\% \\
\hline
BBC03 \& H10 field (0.05$<$z$<$0.5) & 46 & 1 & $<$0.1\% & 4 & 1.7\% & 9 & 3.2\% \\
Bullet \& MS2137 field & 28 & 0 & 1.3\% & 1 & 0.1\% & 1 & $\ll$0.1\% \\
Bullet field & 19 & 0 & 10\% & 1 & 7.1\% & 1 & 0.6\% \\
Bullet field (excl. HLSJ065807--555541) & 18 & 0 & 13\% & 0 & 1.2\% & 0 & $<$0.1\% \\
\enddata

Probabilities reflect the likelihood that the sample is drawn from an identical parent distribution to the Bullet Cluster.
\label{tab:nwarm}
\end{deluxetable}

\subsection{Significance of the warm dust population}
\label{sec:sig}

The number of $L_{\rm IR}$ $<$ 2$\times$10$^{11}$ L$_{\odot}$ sources is small: 37 in the Bullet Cluster, nine in MS2137 and 28 in the two foreground fields (19 of which are in front of the Bullet Cluster). We test whether the lack of galaxies exhibiting warm dust temperatures in MS2137 and the field samples is significant by calculating the likelihood of randomly drawing the observed sample from an identical temperature distribution to the Bullet Cluster members. We confine this analysis to sources with $L_{\rm IR}$ $<$ 2$\times$10$^{11}$ L$_{\odot}$, where the luminosity--temperature relation is flat. Table \ref{tab:nwarm} summarizes the results, concentrating on the probability of selecting the observed number of sources above three temperature thresholds ($T_{\rm dust}$ $=$ 33, 37, 40 K).

The lack of warm dust hosts in MS2137 is not significant. As described in Sections \ref{sec:obs} and \ref{sec:analysis}, there are multiple inconsistencies between the datasets (mid-infrared footprint, 70 \micron{} coverage, depth at 160 \micron{}). Particularly important is the lack of 70 \micron{} coverage (as discussed in Section \ref{sec:valid}), as the peak wavelength of a warm dust component falls below 100 \micron{}. Warm dust sources in MS2137 may simply be misclassified without 70 \micron{} photometry to constrain the peak. Comparing the field samples to the Bullet Cluster, Table \ref{tab:nwarm} reveals a marginally significant difference in the distributions ($>$ 3$\,\sigma$). For $L_{\rm IR}$ $<$ 2$\times$10$^{11}$ L$_{\odot}$, the Bullet Cluster has more $T_{\rm dust}$ $>$ 33 K sources than our foreground fields, and more extreme dust ($T_{\rm dust}$ $>$ 40 K) than the BBC03+H10 sample (also illustrated by the histogram in Figure \ref{fig:lir_temp_blain}). Warm sources are at least preferentially located in the cluster environment.

In summary, the dust temperature distribution of the Bullet Cluster sources is not consistent with the foreground field samples. Here, we formally define `warm dust' sources as those satisfying the following two criteria: (1) $L_{\rm IR}$ $<$ 2$\times$10$^{11}$ L$_{\odot}$, marking the known break in the IR luminosity -- temperature relation; (2) $T_{\rm dust}$ $>$ 37 K, the nearest integer temperature to the 3$\,\sigma$ tail of the BBC03+H10 temperature distribution. There are six warm dust sources in the Bullet Cluster: their SEDs are displayed in the lower two rows of Figure \ref{fig:seds}, and derived properties are also listed in Table \ref{tab:ws}. As dust temperature and redshift are degenerate, it is worth noting that the six warm dust sources all have concurring spectroscopic redshifts from Magellan/IMACS and CTIO/Hydra, ruling out the possibility of foreground interlopers.

\placetable{tab:ws}
\begin{deluxetable}{cccccccc}
\tablecolumns{10}
\tablewidth{0pc}
\tablecaption{Derived properties of the warm dust sources in the Bullet Cluster}
\tablehead{ & \colhead{full ID} & \colhead{$z$} & \colhead{$L_{\rm IR}$ (R09)} & \colhead{SFR (R09)} & \colhead{$T_{\rm dust}$} & \colhead{$M_{\rm dust}$} & \colhead{AGN?} \\
 & & & [$\times$10$^{10}$ $L_\sun$] & [$M_\sun$ yr$^{-1}$] & [K] & [$\times$10$^{8}$ $M_\sun$] & }
\startdata
WD1 & HLSJ065817--560249 & 0.300 & 5.5 $\pm$ 0.3 & 9.5 $\pm$ 0.5 & 39 $\pm$ 1 & 2.1 $\pm$ 0.4 & (1) \\
WD2 & HLSJ065826--555913 & 0.301 & 1.9 $\pm$ 0.2 & 3.3 $\pm$ 0.3 & 43 $\pm$ 3 & 4.4 $\pm$ 0.2 & -- \\
WD3 & HLSJ065829--555316 & 0.304 & 9.4 $\pm$ 0.7 & 16 $\pm$ 1 & 39 $\pm$ 2 & 3.4 $\pm$ 0.6 & -- \\
WD4 & HLSJ065831--560336 & 0.290 & 6.8 $\pm$ 0.4 & 12 $\pm$ 1 & 48 $\pm$ 2 & 1.0 $\pm$ 0.2 & (2) \\
WD5 & HLSJ065842--560041 & 0.296 & 2.5 $\pm$ 0.3 & 4.3 $\pm$ 0.4 & 37 $\pm$ 1 & 1.2 $\pm$ 0.3 & -- \\
WD6 & HLSJ065850--560037 & 0.289 & 4.3 $\pm$ 0.3 & 7.5 $\pm$ 0.5 & 43 $\pm$ 2 & 1.0 $\pm$ 0.2 & -- \\
\enddata

(1) coincident with X-ray point source; see Section \ref{sec:agn} \\
(2) AGN-like $S_{\rm 70}$/$S_{\rm 24}$; right panel, Figure \ref{fig:agn_tests}
\label{tab:ws}
\end{deluxetable}

\subsection{AGN contribution}
\label{sec:agn}

We now explore whether obscured AGN are contributing to the mid-infrared of the warm dust sources, elevating the characteristic dust temperature. We note that AGN could only account for the variation between field and cluster samples if star-forming AGN hosts are more prevalent in clusters. At the redshift of the Bullet Cluster, clusters are likely to contain a lower fraction of AGNs than comparable field samples (a few per cluster; \citealt{mar06-116,gal09-1309}).

We examine several diagnostics to assess AGN contribution to the infrared. First we consult the {\it Chandra} point source catalog for the Bullet Cluster (private communication, M. Markevitch), containing 145 sources in the central $\sim$20' to a flux of 2.5$\times$10$^{-16}$ erg cm$^{-2}$ s$^{-1}$. Six {\it Herschel}-detected, spectroscopic sources are coincident ($\Delta r$ $<$ 1") with {\it Chandra} point sources; three in the foreground and three cluster members. Two of the cluster members have \{$L_{\rm IR}$ $=$2.7$\times$10$^{10}$ L$_{\odot}$, $T_{\rm dust}$ $=$ 27 K\} and \{$L_{\rm IR}$ $=$2.4$\times$10$^{10}$ L$_{\odot}$, $T_{\rm dust}$ $=$ 30 K\} (L$_{\rm X}$ $=$ 4$\times$10$^{42}$ erg s$^{-1}$, 2$\times$10$^{41}$ erg s$^{-1}$ respectively). The third cluster member is a warm dust sources (WD1; f(0.5--2 keV) $=$ 2.83$\times$10$^{-15}$ erg cm$^{-2}$ s$^{-1}$; L$_{\rm X}$ $=$ 8$\times$10$^{41}$ erg s$^{-1}$). Based on X-ray non-detection, the remaining five warm dust sources have X-ray luminosities L$_{\rm X}$ $<$ 7$\times$10$^{40}$ erg s$^{-1}$. Stacking on the optical positions of these five sources in the archival {\it Chandra} X-ray image reveals no significant signal, further constraining the average X-ray luminosity of the other warm dust hosts to L$_{\rm X}$ $<$ 4$\times$10$^{40}$ erg s$^{-1}$.

\begin{figure*}
\centering
\includegraphics[angle=0,scale=0.6]{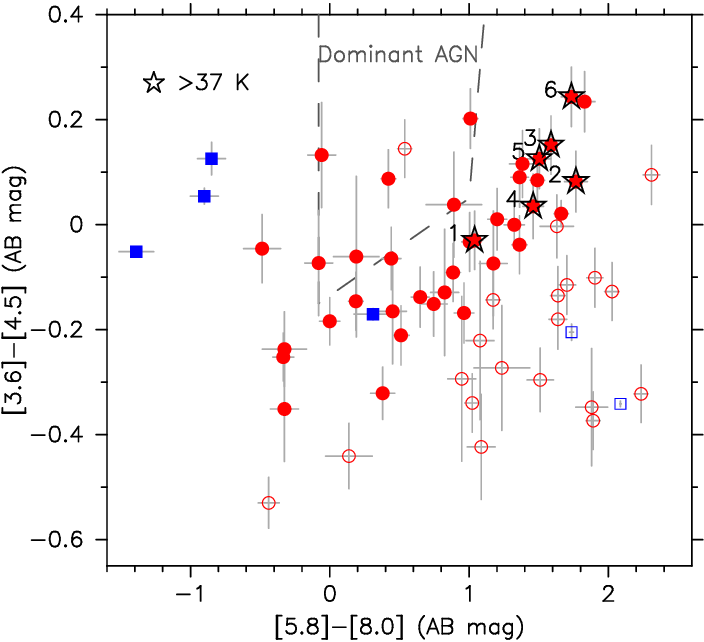}
\hspace{2mm}
\includegraphics[angle=0,scale=0.6]{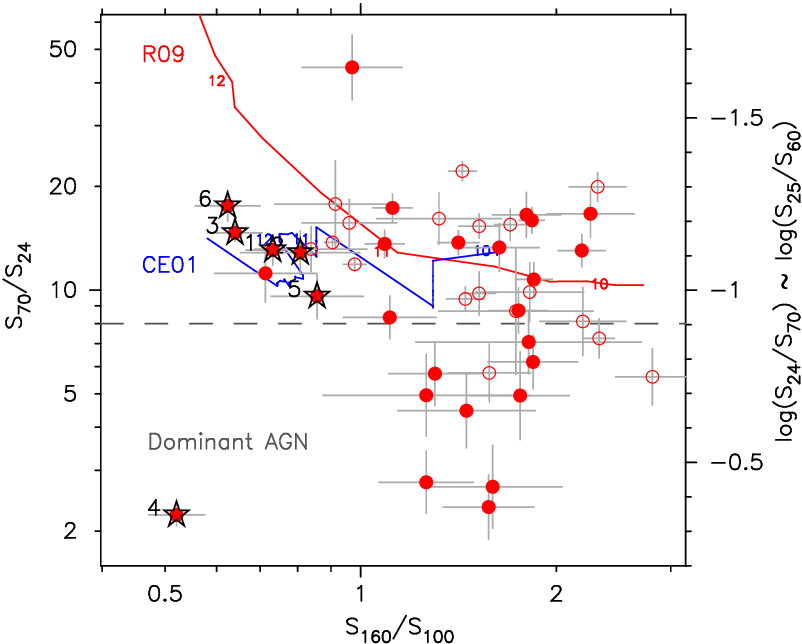}
\caption{Infrared tests for dominant AGN contribution. Bullet Cluster$=$filled red circles, MS2137$=$filled blue squares (foreground fields are in corresponding open symbols). Warm dust sources ($T_{\rm dust}$ $>$ 37 K) are highlighted by black stars and labelled with their WD number for reference (see Table \ref{tab:ws}). {\it Left panel:} IRAC color--color plot with the \citet{ste05-163} AGN criteria marked by grey dashed lines. {\it Right panel:} Mid-IR color--color diagram with the \citet{vei09-628} AGN diagnostic shown by the grey dashed line: sources below have $>$50\% probability of a dominant AGN, and include one warm dust source (WD4). R09 (red) and CE03 (blue) template tracks are shown to indicate the location of normal star-forming galaxies, with log($L_{\rm template}$/$L_{\sun}$)=\{10,11,12\} indicated by small numerals.}
\label{fig:agn_tests}
\end{figure*}

A simple power-law continuum through the IRAC photometry \citep[a simple indicator of an AGN; e.g.][]{don08-111} is not a good fit to the warm dust SEDs. In the left panel of Figure \ref{fig:agn_tests} we show an IRAC color--color diagram for more detail. Following the prescription of \citet{ste05-163}, normal galaxies (without AGN) occupy the region below and to the left of the bound area. Foreground and cluster sources appear separated due to the peak of stellar emission shifting red-ward through the IRAC bands with increasing redshift. Dominant AGN are located within the bound area, and none of the warm dust sources are found here. Galaxies with possible sub-dominant AGN contribution are found to the right of the AGN locus. All of the warm dust sources are found in this region, which may indicate a low level contribution to the mid-infrared, or may be a consequence of an unusual warm dust component. The warm dust source coincident with an X-ray point source (WD1) exhibits the reddest IRAC colors of the warm dust population, and is not consistent with the \citeauthor{ste05-163} AGN criteria.

The right panel of Figure \ref{fig:agn_tests} combines MIPS and PACS mid-IR data as a proxy for the IRAS colors used by \citet{vei09-628} (as in \citealt{hat10-33}). Here, dominant AGN have $S_{\rm 70}$/$S_{\rm 24}$ $<$ 8. CE01 and R09 template tracks indicate approximately where normal star forming-galaxies are located, and five of the six warm dust sources lie in this region (WD1 has $S_{\rm 70}$/$S_{\rm 24}$ $=$ 13.1). WD4 is located below the AGN diagnostic line with $S_{\rm 70}$/$S_{\rm 24}$ $\sim$ 2. However, this source does not resemble the normal AGN population which have much redder $S_{\rm 160}$/$S_{\rm 100}$ colors. Even if WD4 has a significant AGN component, the source does not resemble the typical AGN host population in the mid-infrared.

\subsection{Properties of warm dust sources}

\subsubsection{Spatial distribution}

\begin{figure}
\centering
\includegraphics[angle=270,scale=0.62]{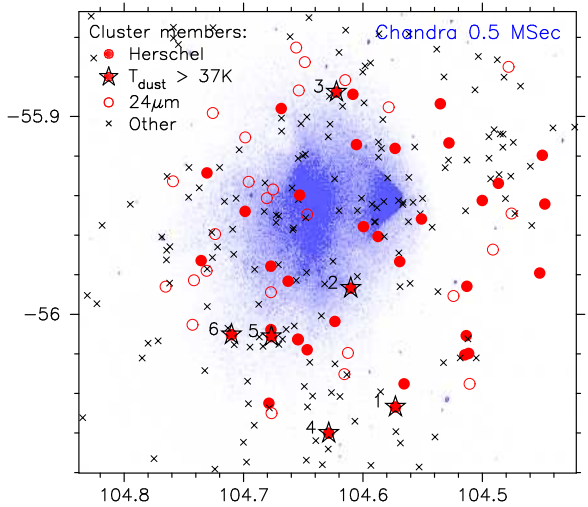}
\caption{Location of the Bullet Cluster members with respect to the diffuse X-ray component detected by {\it Chandra} (blue intensity map). {\it Hershel} detections are shown as filled red circles, with the warm dust sources highlighted by stars and labelled with their WD number for reference.}
\label{fig:xray}
\end{figure}

In terms of simple projected cluster-centric radius, the warm dust sources are similarly distributed to the whole FIR-bright population, i.e. absent from the densest core and found preferentially towards the periphery of the cluster, as illustrated by Figure \ref{fig:xray}. Although spectroscopic fiber placement may contribute, this is consistent with the location of {\it Herschel} cluster sources in clusters with more complete redshift data, e.g. LoCuSS \citep{smi10-18}. Five of the warm dust sources are located to the south of the cluster core, but the velocity offset ranges too widely ($\Delta cz$ $\sim$ 4000 km s$^{-1}$) to be associated with a single substructure (such as an in-falling group or filament). The warm dust sources generally avoid the very cluster center ($>$2--3 arcmin). Although they appear to be located preferentially in the south, this is due to the PACS coverage, which extends to larger radii in this direction (see Figure \ref{fig:footprints}).

The hot, dense intracluster medium (ICM) could plausibly trigger the mechanism responsible for warm dust in some cluster sources e.g. by actively heating dust, or preferentially stripping cooler dust from the outer regions of the galaxy. Figure \ref{fig:xray} displays the diffuse X-ray emission in the Bullet Cluster core, revealing that four of the warm dust sources are located in the vicinity of the densest X-ray gas, in regions where the gas temperature is $>$ 10 KeV (for reference, the X-ray temperature in the core of MS2137 is T$_{\rm X}$ $\sim$ 5--8 KeV; \citealt{cav09-12}). Two warm dust sources are located well away from the brightest X-ray emission (T$_{\rm X}$ $\sim$ 8 KeV), indicating that the extreme X-ray gas is unlikely to play a role. If the dense ICM were responsible for the warm dust, we may expect the phenomenon to be more prevalent in cluster galaxies than the observed $\sim$15\% level (6/37).

\subsubsection{Morphology}

If harassment or major interaction were responsible for the warm dust, the sources may have visible signs of disturbance. On the other hand, an undisturbed morphology would tend to favor a slow process such as strangulation (starvation) by the dense ICM. IMACS imaging (seeing $\sim$0.5 arcsec FWHM) hints at an irregular or disturbed morphology for three of the six warm dust galaxies, although the resolution is too poor for in-depth study of galaxies at z$\sim$0.3. Only one warm dust source (WD4) is covered by high-resolution HST imaging in the core of the Bullet Cluster. \citet{chu10-1536} investigated the source in detail\footnote{Guided by 24 \micron{} flux alone \citet{chu10-1536} classify the galaxy as the only ULIRG candidate in the Bullet Cluster; {\it Herschel} photometry reveals the source to be a sub-LIRG with unusually warm dust and an inflated 24 \micron{} flux compared to $L_{\rm IR}$.} and show it to be a normal barred spiral with no sign of major interaction (their figure 7). The relatively undisturbed morphology suggests that strangulation is the dominant process in, at least, this galaxy \citep{chu10-1536}.

\subsubsection{Dust and stellar mass}

We briefly compare the dust properties to stellar mass, as the mass of the warm dust host galaxy could be an important parameter in stripping or interactions. Stellar mass is determined using the SED-fitting method described in \citet{per08-234}. Dust mass is estimated from the greybody temperature and (restframe) 500 \micron{} flux (calculated from the best fit greybody in Section \ref{sec:fits}), using the formulation of \citet{dra03-241}\footnote{$M_{\rm dust} = (D^2 f_{\rm 500})/(\kappa_{\rm abs} B_{\lambda}(T_{\rm dust}))$, where $\kappa_{\rm abs}$(500 \micron{}) $=$ 0.95 cm$^2$ g$^{-1}$}. Dust masses are estimated from 500 \micron{} to minimize the dependence on temperature. We note that the 500 \micron{} flux may underestimate the total mass by missing the warmest dust (which contributes only a fraction of the emission at this wavelength). Furthermore, dust mass estimates from the \citeauthor{dra03-241} models are wavelength dependent; dust mass derived from the flux at a rest wavelength of 200 \micron{} is systematically lower by $\sim$0.2 dex, although the distribution of dust masses remains the same.

\begin{figure}
\centering
\includegraphics[angle=270,scale=0.73]{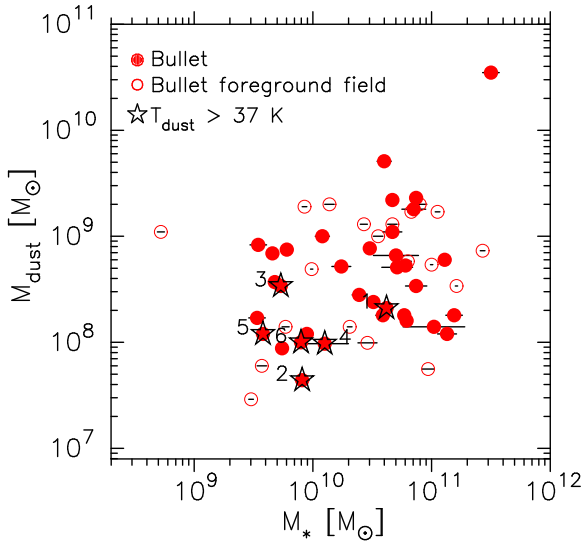}
\caption{Dust mass (calculated from the greybody fit parameters) versus stellar mass (from a full optical/NIR SED fit) for the Bullet Cluster (filled red circles) and foreground field (open red circles) sources detected by {\it Herschel}. The warm dust sources are highlighted by stars as in previous plots, and tend towards the lower end of the observed mass distribution.}
\label{fig:mass}
\end{figure}

Figure \ref{fig:mass} plots dust mass against stellar mass, showing that the warm dust hosts tend to be amongst the least massive of our observed star-forming galaxies (4/6 have $M_*$ $<$ 10$^{10}$ M$_{\sun}$). They are also all towards the lower end of the dust mass distribution, suggesting that the high characteristic temperature is not due to an influx of additional warm dust, but rather the conversion or removal of cooler dust.

\subsubsection{Interpretation of dust component SED}

Resolved infrared studies of nearby galaxies have investigated the internal gradients of the dust properties. Although the sample sizes are small, the general consensus is that bulges of late-type galaxies harbor enhanced dust heating \citep{alt98-807,mel02-709}. Recently, \citet{eng10-56} explored the bulges and discs of 13 nearby galaxies with {\it Herschel}, and confirmed the general trend of warmer centers compared to the outskirts. Stripping preferentially removes outer material, so the inferred characteristic temperature of an unresolved source would be seen to rise, as the cooler dust is removed first. The truncation of gas disks in cluster galaxies has long been observed (e.g. \citealt{cay94-1003}), while the suspected, associated truncation of the dust disk has recently been confirmed using {\it Herschel} data by \citet{cor10-49}. The relatively low dust mass estimates of the warm dust sources in our study appears to support such a scenario.

\begin{figure}
\centering
\includegraphics[angle=270,scale=0.65]{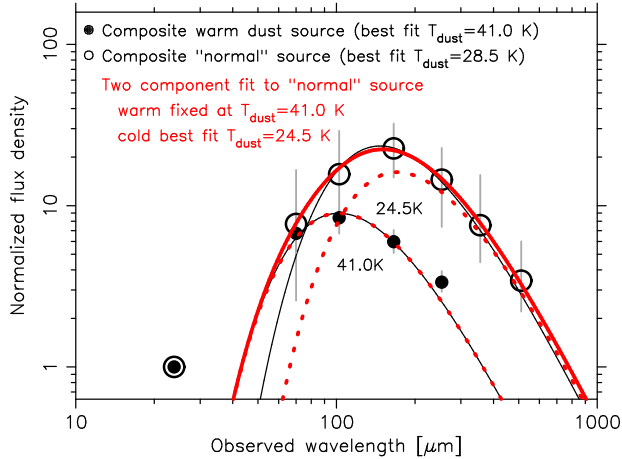}
\caption{Composite SEDs normalized at 24 \micron{} for the Bullet Cluster warm dust sources ($T_{\rm dust}$ $>$ 37 K), shown by solid circles, and `normal' star-forming cluster galaxies, as open circles. Note that the 24 \micron{} points for the two SEDs are coincident, due to the normalization. The range within each population is indicated by the grey bars. The best fitting greybodies are shown by thin black lines ($T_{\rm dust}$ $=$ 41 K and 28.5 K respectively). The normal star-forming composite SED is also fit by a two component model, each shown by thick red dotted lines (total as red solid line): a warm component identical to the best fit warm dust source greybody, and a colder dust component (derived to have $T_{\rm dust}$ $=$ 24.5 K). The dust in the lower temperature component would need to be removed to transform a normal star-forming cluster galaxy into a warm dust source.}
\label{fig:composite}
\end{figure}

We investigate this further by examining whether the SED shape of warm dust sources can be recovered via the removal of cold dust from a `normal' star-forming cluster galaxy. For the warm dust population and, separately, the remaining Bullet Cluster members, a composite SED is derived by summing the individual SEDs after normalizing at 24 \micron{}. At the redshift of the Bullet Cluster, 24 \micron{} traces the embedded star formation, which is most likely to remain unchanged by a stripping process. Figure \ref{fig:composite} shows the resulting greybody fits to the two composite SEDs, which recovers the population average parameters $T_{\rm dust}$ $=$ 41 K and $T_{\rm dust}$ $=$ 28.5 K respectively. For a given 24 \micron{} flux, the warm dust sources have a smaller infrared dust component (and therefore lower dust mass) than normal star-forming cluster galaxies (as already observed in Figure \ref{fig:mass}). The composite SEDs also suggest that an infrared luminosity extrapolated from 24 \micron{} would generally overestimate the true $L_{\rm IR}$ for the warm dust sources, a result confirmed by Figure \ref{fig:24um_comp}, in Appendix \ref{app:24um}.

Assuming that the average warm dust source can be obtained by removing dust from the average `normal' star-forming galaxy, the stripped material equals the difference between the two composite SEDs. Subtracting the 41 K greybody from the other composite SED, leaves a 24.5 K cold component (demonstrated in Figure \ref{fig:composite}). Using a typical 24 \micron{} flux (median of our {\it Herschel}-detected sample, $S_{24}$ $=$ 0.3 mJy), the dust mass of this cold component would be M$_{\rm dust,cold}$ $\sim$ 8$\times$10$^7$ M$_{\sun}$. The warm dust sources have a mean dust mass of $\sim$5$\times$10$^7$--5$\times$10$^8$ M$_{\sun}$, indicating that in this scenario, $\sim$30--50\% of the progenitor dust mass was stripped. Adding this amount of cold dust back into each of the individual warm dust sources, gives initial $T_{\rm dust}$ $\sim$ 28 K and $L_{\rm IR}$ $\sim$ 5$\times$10$^{10}$ -- 2$\times$10$^{11}$ L$_{\odot}$, which would sit comfortably amongst the normal star-forming galaxy population within the cluster (see Figure \ref{fig:lir_temp_blain} for comparison). These progenitors would also have dust masses towards the upper envelope of the normal star-forming galaxies (M$_{\rm dust}$ $\sim$ 8$\times$10$^{8}$ -- 8$\times$10$^{9}$ M$_{\sun}$; see Figure \ref{fig:mass}). It is not surprising that the detected warm dust sources are the progeny of galaxies lying at the high luminosity end of the observed range; less IR-luminous galaxies undergoing the same stripping process would have enough dust removed to render them undetectable in our {\it Herschel} imaging ($L_{\rm IR}$ $<$ 10$^{10}$ L$_{\odot}$). Stacking on {\it Herschel}-undetected 24 \micron{} cluster sources recovers a marginal PACS detection at 70 \micron{} alone. Without a deeper FIR stack, we cannot confirm the presence of a lower-$L_{\rm IR}$, warm dust population.

The stripping scenario appears plausible from the viewpoint of the total dust mass budget, viability of the progenitors, and the preferential stripping of lower temperature gas. However, the observed central dust temperature of nearby field galaxies is not as warm ($\la$30 K) as the characteristic dust in some cluster sources ($\ga$40 K). The high dust temperature observed in several cluster galaxies appears to require additional heating, which may also be a consequence of the stripping mechanism (few of the resolved local field galaxies will have experience such a process). As a galaxy interacts with the ICM, central star formation may be triggered, in turn re-heating the surrounding dust. Stellar masses indicate that warm dust is located in less massive hosts, which suggests that molecular clouds in low mass galaxies are shocked and compressed as they encounter the dense ICM, triggering star formation in a relatively small volume, which may result in unusually warm dust \citep{bek03-13}. Heating and removal of outer (cold) dust are not exclusive, with both plausible components of a stripping mechanism that acts to increase the characteristic dust temperature of a spatially-unresolved source.

\section{Summary}
\label{sec:summary}

We explore far-infrared bright cluster members from two dense environments at z$\sim$0.3, covered by the ``{\it Herschel} Lensing Survey": the merging Bullet Cluster and the more relaxed MS2137. At this redshift, characteristic dust temperature $T_{\rm dust}$ is well correlated with the PACS color $S_{\rm 160}$/$S_{\rm 100}$, which straddles the FIR component peak. Previous studies have shown that sub-LIRG galaxies display dust temperatures $\sim$30 K, while dust in more luminous sources is warmer, $T_{\rm dust}$ $=$ 30--70 K.

\begin{itemize}
\item Several warm dust galaxies are discovered in the Bullet Cluster, having higher than expected characteristic dust temperature given their IR luminosity ($T_{\rm dust}$ $>$37 K at $L_{\rm IR}$ $<$ 10$^{11}$ L$_{\odot}$).

\item No warm dust galaxies are found in MS2137, although this could be an observational bias due to the lack of 70 \micron{} coverage or the smaller FIR field-of-view.

\item Two field samples -- the foreground of the Bullet observations, with identical photometric data to the cluster, and a larger literature sample (\citealt{bla03-733,hwa10-75}) -- are not compatible with the Bullet Cluster population, suggesting that warm dust galaxies are, at least, preferentially located in over-dense environments.
\end{itemize}

We examine scenarios which could be responsible for the unusual population in the Bullet Cluster. One source is coincident with an X-ray point source (L$_{\rm x}$ $=$ 8$\times$10$^{41}$ erg s$^{-1}$), but the remainder show no X-ray signature or optical/near-infrared line emission indicative of an AGN. The $S_{\rm 70}$/$S_{\rm 24}$ color of a second warm dust source suggests a substantial mid-infrared AGN contribution, although the SED does not resemble a typical AGN-host.

Composite SEDs for a warm dust galaxy and a `normal' star-forming cluster galaxy are derived to explore a simple dust stripping scenario, in which dust is preferentially removed from the outskirts of a galaxy where dust temperature is lower. The mass of the removed dust is plausible, requiring 30--50\% stripped from a normal star-forming galaxy to leave an SED similar to the warm dust sources. The central regions of (spatially-resolved) nearby galaxies are cooler than the characteristic temperature of the warm dust sources, suggesting that heating may be needed in addition to cool dust removal, a possible consequence of the same stripping mechanism. The only warm dust source with high-resolution imaging for morphological analysis shows no sign of disturbance, favoring a slow stripping mechanism such as strangulation by the ICM (rather than e.g. harassment). However, we find no correlation between the location of warm dust sources and the densest ICM, indicating that the process may be triggered by the relatively less dense environment in the cluster outskirts, and suggesting that such sources should be observable in other, less massive clusters.

With analysis of the full {\it Herschel} cluster sample available from the lensing surveys, we will be able to better constrain the fraction of warm dust sources in the dense environment, and understand any dependence on environment, particularly the local ICM density, and host properties such as stellar mass and morphology.

\acknowledgments

This work is partially based on observations made with the {\it Herschel} Space Observatory, a European Space Agency Cornerstone Mission with significant participation by NASA. Support for this work was provided by NASA through an award issued by JPL/Caltech. We would also like to thank the HSC and NHSC consortia for support with data reduction. This work has made use of the private version of the Rainbow Cosmological Surveys Database, which is operated by the Universidad Complutense de Madrid (UCM). We would also like to thank Maxim Markevitch for providing the Bullet Cluster X-ray point source catalog, and Jean-Gabriel Cuby for the VLT/HAWK-I images.

\appendix
\section{$L_{\rm IR}$ extrapolated from 24 \micron{}}
\label{app:24um}

\begin{figure}
\centering
\includegraphics[angle=270,scale=0.9]{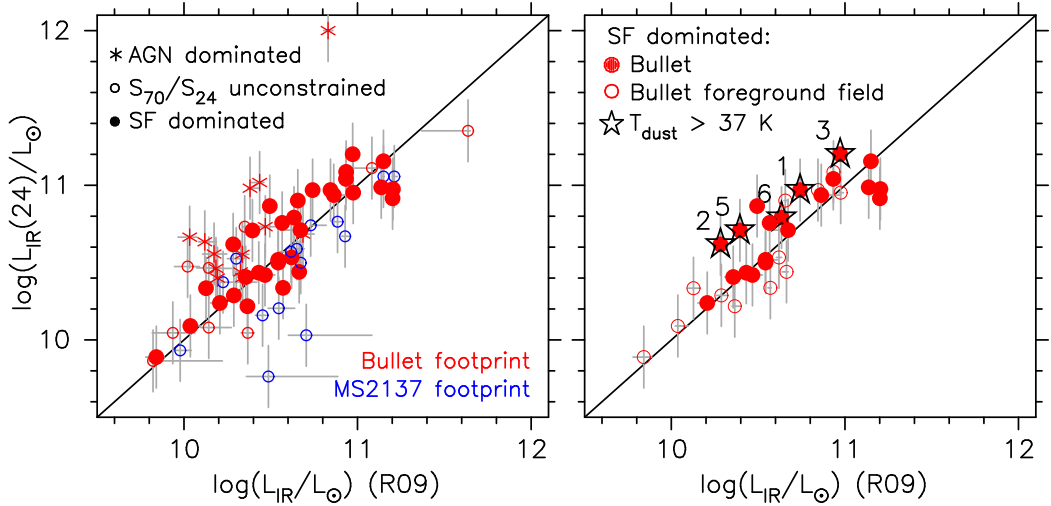}
\caption{Comparison between 24 \micron{}-extrapolated $L_{\rm IR}$(24) and the {\it Herschel} $L_{\rm IR}$ (via R09 templates). {\it Left panel:} Sources in the Bullet Cluster observations (both field and cluster members) are in red, MS2137 are in blue. The total rms scatter is 0.30 dex. Stars identify dominant AGN hosts (from Figure \ref{fig:agn_tests}, right panel), a component unaccounted for in the simple 24 \micron{} extrapolation. Excluding these, the rms scatter is reduced to 0.23 dex. The significance of any AGN component in sources without 70 \micron{} photometry is unknown (open circles). The remaining sources can be considered star-formation dominated (filled circles). {\it Right panel:} The star-formation dominated sources replotted for clarity, with Bullet Cluster$=$red filled circles, MS2137$=$blue filled squares and field samples in corresponding open symbols. The rms scatter in this population is 0.17 dex. The warm dust sources all have over-estimated $L_{\rm IR}$(24). Note that WD4 is the AGN host at the very top of the left panel, so does not appear in the right panel.}
\label{fig:24um_comp}
\end{figure}

Prior to the launch of {\it Herschel}, total infrared luminosity was often extrapolated from 24 \micron{} photometry alone, where the mid-infrared sensitivity of {\it Spitzer} peaked. Initial {\it Herschel} observations show that such 24 \micron{}-derived IR-luminosities are reasonably reliable for star-formation dominated galaxies at low redshift (average scatter of 0.15 dex for sources at z $<$ 1.5; \citealt{elb10-29}). We verify this for cluster sources by calculating SFR$_{\rm 24}$ via equation 14 from \citet{rie09-556}, and use the \citet{ken98-189} formula\footnote{SFR$_{\rm IR}$ [$M_{\sun}$ yr$^{-1}$] $=$ 4.5$\times$10$^{-44}$ $L_{\rm IR}$ [erg s$^{-1}$]} to convert back to IR luminosity. After removing sources suspected of a dominant AGN contribution (those below the line in the right-hand panel of Figure \ref{fig:agn_tests}), which are unaccounted for in the templates, $L_{\rm IR}$(24) and $L_{\rm IR}$({\it Herschel}) correlate reasonably well for our galaxies (rms scatter of 0.23 dex, Figure \ref{fig:24um_comp}). It is difficult to constrain the AGN contribution for sources without 70 \micron{} photometry. Including only confirmed star formation-dominated sources (i.e. excluding those without 70 \micron{} data), the rms scatter decreases further to 0.17 dex. The warm dust sources discovered in this paper (highlighted by squares in Figure \ref{fig:24um_comp}), all have an over-predicted $L_{\rm IR}$ from 24 \micron{}. This suggests a mechanism that modifies the cold dust component while leaving the mid-infrared, which traces embedded star formation, unchanged (Figure \ref{fig:composite}).

\section{Exploring the luminosity discrepancy}
\label{app:lum}

\begin{figure*}
\centering
\includegraphics[angle=270,scale=0.9]{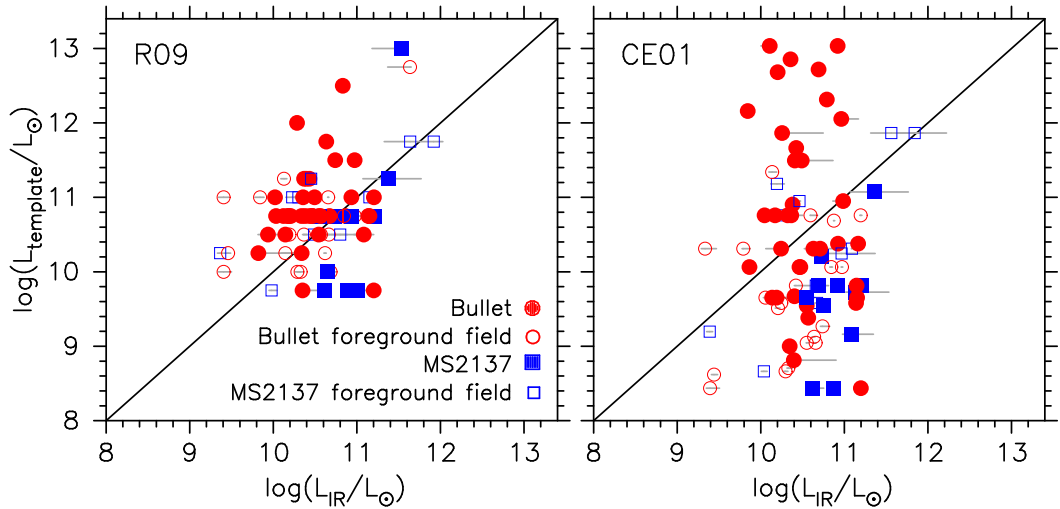}
\caption{$L_{\rm template}$ compared to the integrated $L_{\rm IR}$ of the normalized best-fitting template (see text for details): left$=$R09, right$=$CE01. Bullet Cluster$=$red filled circles, MS2137$=$blue filled squares. Field samples in corresponding open symbols. Note that the smaller apparent scatter for the R09 set is artificial as no templates exist log($L_{\rm template}$) $<$ 9.75.}
\label{fig:int_temp_comp}
\end{figure*}

For several Bullet Cluster galaxies, we identified a discrepancy between the nominal luminosity class of the best-fitting local template and the integrated luminosity of that template when normalized to the observations (i.e. $L_{\rm IR}$ $\ne$ $L_{\rm template}$). In \citet{rex10-13} we discovered a similar (but opposite) discrepancy for high redshift sources, such that the shape of the SED resembles a low luminosity local galaxies (perhaps because star formation in local galaxies is more concentrated; i.e. \citealt{ruj11-tmp}). Large blank-field surveys have not reported such a mismatch (up to $z$ $\sim$ 1.5; e.g. \citealt{elb10-29}), although this may be because an automated routine would report a good statistical fit, regardless of the nominal template luminosity class. The cluster sample presented in this paper displays a large range in this template mis-match, with no obvious correlation between $L_{\rm template}$ and $L_{\rm IR}$ (Figures \ref{fig:dusttemp} and \ref{fig:int_temp_comp}). Note that the apparently smaller scatter for the R09 templates than CE03 is merely a consequence of a more limited range in $L_{\rm template}$: 9.75 $\le$ log($L_{\rm template}$) $\le$ 13.0 compared to 8.4 $\le$ log($L_{\rm template}$) $\le$ 13.5. 

\begin{figure}
\centering
\includegraphics[angle=0,scale=0.9]{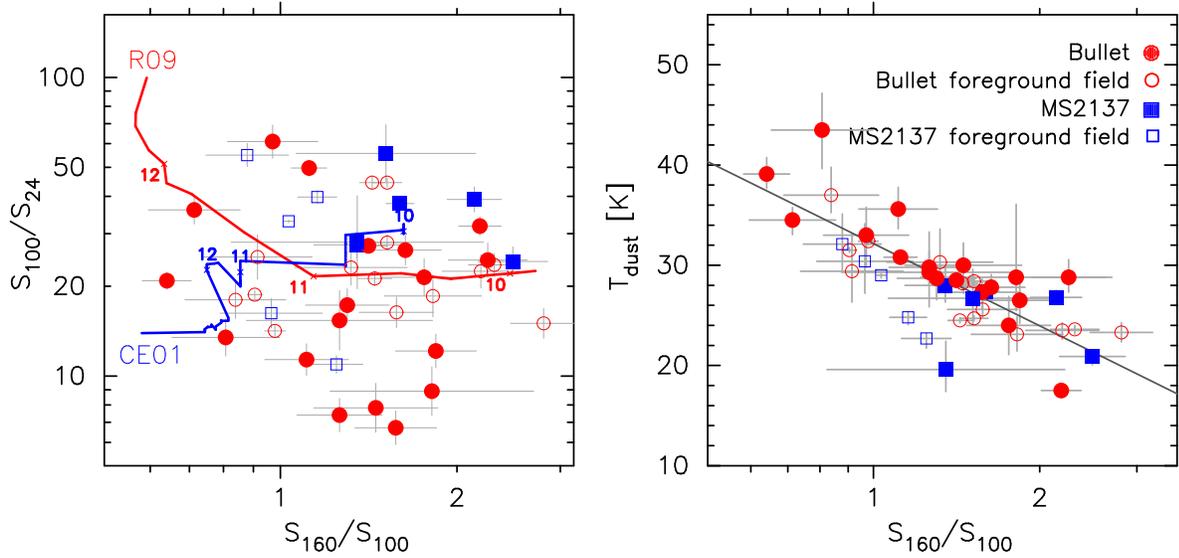}
\caption{PACS color S$_{160}$/S$_{100}$ plotted against S$_{100}$/S$_{24}$ ({\it left panel}) and $T_{\rm dust}$ ({\it right panel}). Symbols and colors as in the previous plot. In the left panel, tracks of increasing luminosity are shown for the R09 (red) and CE01 (blue) templates at a redshift of z$=$0.3. The track positions for log($L_{\rm template}$)$=$\{10,11,12\} are indicated by small numerals. The right panel shows that S$_{160}$/S$_{100}$ is a reasonable probe of $T_{\rm dust}$ at this redshift (rms dispersion in temperature of 3.7 K).}
\label{fig:pacscols}
\end{figure}

\begin{figure*}
\centering
\includegraphics[angle=0,scale=0.93]{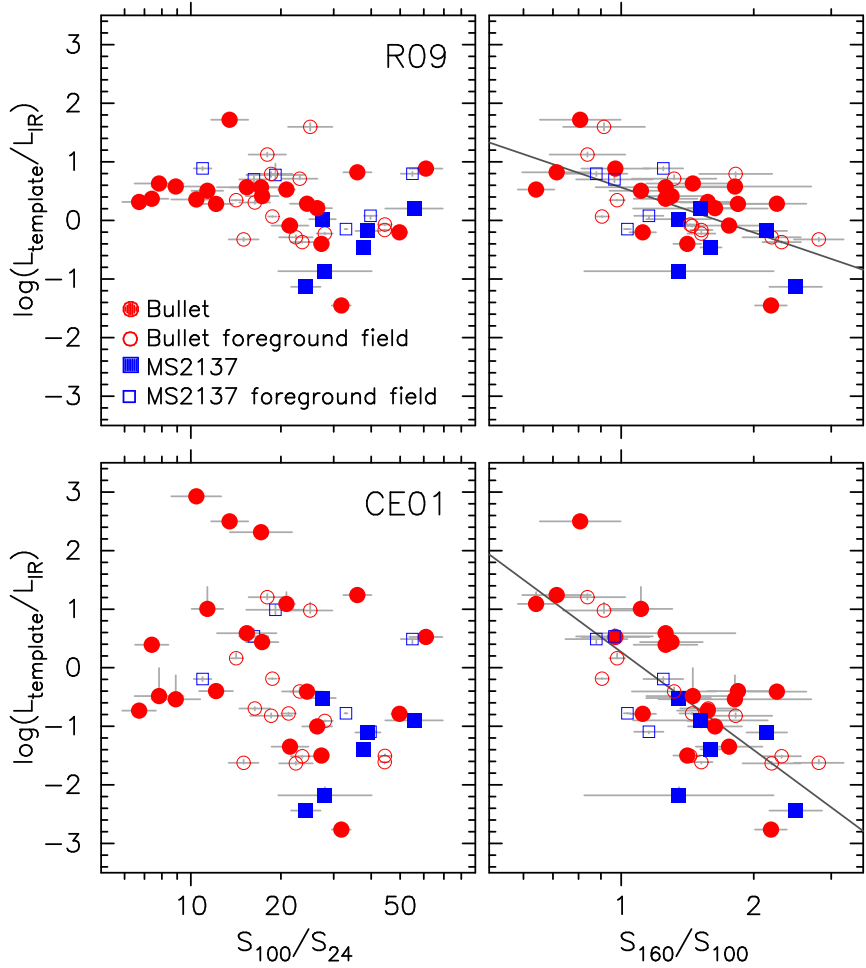}
\caption{The template mismatch, log($L_{\rm template}$/$L_{\rm IR}$) (upper row$=$R09; lower row$=$CE01) compared to $S_{\rm 100}$/$S_{\rm 24}$ ({\it left panels}) and $S_{\rm 160}$/$S_{\rm 100}$ ({\it right panels}). The analogous plots for characteristic dust temperature ($T_{\rm dust}$) are given in Figure \ref{fig:dusttemp}. Symbols and colors as in the previous plots. The discrepancy between $L_{\rm template}$ and $L_{\rm IR}$ is correlated with the FIR component peak, traced by $S_{\rm 160}$/$S_{\rm 100}$ (at z$\sim$0.3) and driven in part by dust temperature.}
\label{fig:lfir_mismatch}
\end{figure*}

Initial analysis of far-infrared data from HLS and the Local Cluster Substructure Survey (LoCuSS) suggested a large fraction of cluster galaxy SEDs were not well fit by local templates of comparable luminosity \citep{raw10-14,per10-40,smi10-18}. The templates predict S$_{100}$/S$_{24}$ $\sim$ 20--30 for $L_{\rm IR}$ $<$ 10$^{11}$ galaxies at z$\sim$0.3 (Figure \ref{fig:pacscols}, left panel), whereas the cluster sources show a wide variety of colors (5 $<$ S$_{100}$/S$_{24}$ $<$ 60). Although a low S$_{100}$/S$_{24}$ color can be attributed to an AGN increasing the flux at 24 \micron{}, larger ratios are harder to explain. These preliminary studies asserted that 25--50\% of the star-forming cluster population had larger than expected S$_{100}$/S$_{24}$ (dubbed `100 \micron{} excess'). However, S$_{100}$/S$_{24}$ tracks predicted by R09 and CE01 move in opposite directions with increasing luminosity, illustrating our uncertainty in this color space. The difference results from an elevated dust continuum level for higher luminosity templates in CE01, coupled to the deeper silicate absorption feature for higher luminosity templates in R09. The colors from this current study show a similar scatter around the template tracks (25\% have S$_{100}$/S$_{24}$ $>$ 30; Figure \ref{fig:pacscols}) but there is no general correlation of S$_{100}$/S$_{24}$ with the luminosity discrepancy (Figure \ref{fig:lfir_mismatch}, left panels). The scatter in S$_{100}$/S$_{24}$ is more likely to reflect different levels of strong silicate absorption and/or PAH features in the 24 \micron{} band. S$_{100}$/S$_{24}$ is a poor probe of the template discrepancy.

In contrast, both template sets display the general trend that increasing luminosity decreases S$_{160}$/S$_{100}$ (Figure \ref{fig:pacscols}, left panel). This is simply due to the FIR peak shifting blue-wards through the 160 and 100 \micron{} bands, for sources at z$\sim$0.3. The right panels of Figure \ref{fig:lfir_mismatch} show that as the peak shifts further from the mean wavelength (corresponding to S$_{160}$/S$_{100}$$\sim$1.2), the mismatch between $L_{\rm template}$ and $L_{\rm IR}$ becomes more pronounced. Figure \ref{fig:dusttemp} shows that the template discrepancy in our sample is highly correlated with $T_{\rm dust}$ (as derived from a greybody). $S_{\rm 160}$/$S_{\rm 100}$ is a useful tracer of $T_{\rm dust}$ for galaxies at this redshift (Figure \ref{fig:pacscols}, right panel), and the correlation with template discrepancy is a consequence of the dependency on $T_{\rm dust}$.

\end{document}